%% file: main.tex
\pdfoutput=1
\documentclass{article}

\usepackage{PRIMEarxiv}

\usepackage[utf8]{inputenc}
\usepackage[T1]{fontenc}
\usepackage{amssymb}
\usepackage{amsmath}
\usepackage{amsfonts}
\usepackage{booktabs}
\usepackage{multirow}
\usepackage{array}
\usepackage{tabularx}
\usepackage{float}
\usepackage{placeins}
\usepackage{nicefrac}
\usepackage{microtype}
\usepackage{graphicx}
\graphicspath{{media/}}
\usepackage{xcolor}
\usepackage{fancyhdr}
\usepackage[numbers,sort&compress]{natbib}
\usepackage{url}
\usepackage{hyperref}
\hypersetup{colorlinks=true, linkcolor=blue, citecolor=blue, urlcolor=blue}
\hypersetup{pdftitle={Crowds Believe in You: Online Prediction Markets Overestimate How Often People Achieve Their Own Goals},
  pdfauthor={Zihan Chen, Di Zhu, Lei Nico Zheng},
  pdfkeywords={prediction markets, wisdom of crowds, planning fallacy, self-prediction, social prediction, forecast calibration, online communities},
  pdfsubject={Calibration of prediction markets about people's own goals on Manifold}}

\input{generated/numbers}

\title{Crowds Believe in You:\\Online Prediction Markets Overestimate\\How Often People Achieve Their Own Goals}

\author{
  Zihan Chen \\
  Stevens Institute of Technology \\
  Hoboken, NJ, USA \\
  \texttt{zchen61@stevens.edu} \\
  \And
  Di Zhu \\
  Stevens Institute of Technology \\
  Hoboken, NJ, USA \\
  \texttt{dzhu1@stevens.edu} \\
  \And
  Lei Nico Zheng \\
  University of Massachusetts Boston \\
  Boston, MA, USA \\
  \texttt{lei.zheng@umb.edu} \\
}

\begin{document}
\maketitle

\input{sections/00_abstract}

\keywords{prediction markets \and wisdom of crowds \and planning fallacy \and self-prediction \and social prediction \and forecast calibration \and online communities}

\input{sections/01_introduction}
\input{sections/02_background}
\input{sections/03_method}
\input{sections/04_results}

\input{sections/05_discussion}
\input{sections/06_conclusion}

\bibliographystyle{unsrtnat}
\bibliography{references}

\appendix
\input{sections/A_appendix}

\end{document}

%% file: generated/numbers.tex
\newcommand{\BndExtAsPriced}{$-$0.9}
\newcommand{\BndSelfAsPriced}{$-$4.1}
\newcommand{\CensusFirstYear}{2021}
\newcommand{\CensusLastYear}{2026}
\newcommand{\CrePosYesB}{0.18}
\newcommand{\CrePosYesP}{$p=.088$}
\newcommand{\CrePosYesSE}{0.11}
\newcommand{\DayExtCreatorUpAbs}{1.0}
\newcommand{\DayExtObsTowardAbs}{9.6}
\newcommand{\DayExtObsUpAbs}{4.5}
\newcommand{\DaySelfCreatorTowardAbs}{2.2}
\newcommand{\DaySelfCreatorUpAbs}{2.9}
\newcommand{\DaySelfObsTowardAbs}{4.6}
\newcommand{\DecExtCreatorAbs}{3.5}
\newcommand{\DecExtCreatorShare}{9\%}
\newcommand{\DecExtObsAbs}{34.0}
\newcommand{\DecExtTotalAbs}{37.3}
\newcommand{\DecSelfAuto}{$-$1.5}
\newcommand{\DecSelfAutoSE}{0.2}
\newcommand{\DecSelfCreatorAbs}{10.6}
\newcommand{\DecSelfCreatorLateShare}{47\%}
\newcommand{\DecSelfCreatorShare}{38\%}
\newcommand{\DecSelfObsAbs}{18.7}
\newcommand{\DecSelfTotalAbs}{27.8}
\newcommand{\DiffGapFinal}{$-$3.5}
\newcommand{\DiffGapFinalP}{$p<.001$}
\newcommand{\DiffGapFinalSE}{0.6}
\newcommand{\DiffGapMidAbs}{4.8}
\newcommand{\DiffGapThreeAbs}{2.2}
\newcommand{\DiffRangeMidGap}{$-$8.2}
\newcommand{\DiffSameGapAday}{$-$1.8}
\newcommand{\DiffSameGapThree}{$-$0.8}
\newcommand{\EvContinuity}{97.2\%}
\newcommand{\EvContinuityOld}{85.9\%}
\newcommand{\EvFinalMatch}{99.4\%}
\newcommand{\EvFinalMatchOld}{91.9\%}
\newcommand{\EvLateFills}{37,578}
\newcommand{\EvPurchasesCancelledShare}{17\%}
\newcommand{\ExtBrier}{0.049}
\newcommand{\ExtCalInt}{$-$0.14}
\newcommand{\ExtCalIntSE}{0.05}
\newcommand{\ExtCap}{five}
\newcommand{\ExtDefaultOpen}{90\%}
\newcommand{\ExtGapFinal}{$-$0.9}
\newcommand{\ExtGapFinalAbs}{0.9}
\newcommand{\ExtGapFinalSE}{0.3}
\newcommand{\ExtGapOpen}{$-$7.6}
\newcommand{\ExtRangeMidGap}{$-$1.2}
\newcommand{\ExtRangeMidGapSE}{1.6}
\newcommand{\ExtReliability}{0.001}
\newcommand{\ExtResolution}{0.195}
\newcommand{\ExtYesAtFifty}{47\%}
\newcommand{\FdDiffUpAbs}{3.6}
\newcommand{\FdDiffUpCtrlAbs}{3.0}
\newcommand{\FdDiffUpCtrlSE}{0.6}
\newcommand{\FdDiffUpP}{$p<.001$}
\newcommand{\FdDiffUpSE}{0.6}
\newcommand{\FdNeverDiffUp}{$+$5.0}
\newcommand{\FdNeverDiffUpP}{$p<.001$}
\newcommand{\FdNeverDiffUpSE}{0.9}
\newcommand{\FdNeverExtUpAbs}{4.3}
\newcommand{\FdNeverExtYes}{37\%}
\newcommand{\FdNeverSelfWantedUpAbs}{0.7}
\newcommand{\FdNeverSelfWantedUpSE}{0.5}
\newcommand{\FdNeverSelfWantedYes}{38\%}
\newcommand{\FdSelfUnwantedUpAbs}{6.4}
\newcommand{\FdSelfUnwantedUpSE}{2.1}
\newcommand{\FdSelfWantedUpAbs}{0.8}
\newcommand{\FdSelfWantedUpSE}{0.5}
\newcommand{\FdSelfWantedYes}{44\%}
\newcommand{\GoalEffortGap}{$-$6.3}
\newcommand{\GoalEffortGapSE}{1.0}
\newcommand{\GoalOutcomeGap}{$-$3.3}
\newcommand{\GoalOutcomeGapSE}{0.9}
\newcommand{\GoalReg}{$-$2.4}
\newcommand{\GoalRegHolm}{$p=.325$}
\newcommand{\GoalRegP}{$p=.054$}
\newcommand{\GoalRegSE}{1.2}
\newcommand{\GoalUnwantedGap}{$-$0.5}
\newcommand{\GoalUnwantedGapHi}{$+$2.4}
\newcommand{\GoalUnwantedGapLo}{$-$3.4}
\newcommand{\GoalUnwantedGapSE}{1.5}
\newcommand{\GoalUnwantedN}{294}
\newcommand{\GoalUnwantedYes}{27\%}
\newcommand{\GoalWantedGap}{$-$4.6}
\newcommand{\GoalWantedGapSE}{0.5}
\newcommand{\HistReg}{$-$0.5}
\newcommand{\HistRegSE}{0.3}
\newcommand{\HistThirtyMidGap}{$-$13.7}
\newcommand{\HistZeroMidGap}{$-$6.6}
\newcommand{\HoAwareSelfDiff}{$-$0.0021}
\newcommand{\HoAwareSelfHi}{0.0022}
\newcommand{\HoAwareSelfLo}{$-$0.0058}
\newcommand{\HoAwareSelfRel}{1.5\%}
\newcommand{\HoCDAwareSelfRel}{3.5\%}
\newcommand{\HoDayPooledSelfHi}{$-$0.0008}
\newcommand{\HoDayPooledSelfLo}{$-$0.0114}
\newcommand{\HoDayPooledSelfRel}{2.5\%}
\newcommand{\LmDayBeforeDiff}{$-$4.6}
\newcommand{\LmDayBeforeDiffSE}{0.8}
\newcommand{\LmDayBeforeExtYesFifty}{50\%}
\newcommand{\LmDayDiff}{$-$2.4}
\newcommand{\LmDayDiffP}{$p=.021$}
\newcommand{\LmDayDiffSE}{1.0}
\newcommand{\LmDayNeverDiff}{$-$5.5}
\newcommand{\LmDayNeverDiffSE}{1.7}
\newcommand{\LmFinalDiff}{$-$2.4}
\newcommand{\LmHourDiff}{$-$1.1}
\newcommand{\LmHourDiffSE}{1.1}
\newcommand{\LmHourExtYesFifty}{43\%}
\newcommand{\LmHourNeverDiff}{$-$3.9}
\newcommand{\LmHourNeverDiffSE}{1.7}
\newcommand{\LmOpenDiffAbs}{3.0}
\newcommand{\LmOpenDiffSE}{1.3}
\newcommand{\LmOpenNeverDiff}{$-$1.2}
\newcommand{\LmOpenNeverDiffSE}{1.8}
\newcommand{\LmOpenSelfYesFifty}{45\%}
\newcommand{\LmWeekBeforeDiff}{$-$5.3}
\newcommand{\LmWeekBeforeDiffSE}{0.9}
\newcommand{\LmWeekBeforeExtYesFifty}{48\%}
\newcommand{\LmWeekBeforeNeverDiff}{$-$7.6}
\newcommand{\LmWeekBeforeNeverDiffSE}{1.5}
\newcommand{\LmWeekBeforeSelfYesFifty}{39\%}
\newcommand{\LmWeekDiff}{$-$2.9}
\newcommand{\LmWeekDiffP}{$p=.008$}
\newcommand{\LmWeekDiffSE}{1.1}
\newcommand{\LmWeekExtYesFifty}{41\%}
\newcommand{\LmWeekSelfYesFifty}{38\%}
\newcommand{\NCensus}{194,347}
\newcommand{\NExt}{6,272}
\newcommand{\NExtCreators}{2,624}
\newcommand{\NSame}{4,486}
\newcommand{\NSameCreators}{1,222}
\newcommand{\NSelf}{6,150}
\newcommand{\NSelfCreators}{1,660}
\newcommand{\NSelfDropped}{120}
\newcommand{\NeverDiffAbs}{5.7}
\newcommand{\NeverDiffP}{$p<.001$}
\newcommand{\NeverDiffSE}{1.0}
\newcommand{\RegFive}{$-$1.9}
\newcommand{\RegFiveAbs}{1.9}
\newcommand{\RegFiveMarketsSelf}{5,470}
\newcommand{\RegFiveP}{$p=.006$}
\newcommand{\RegFiveSE}{0.7}
\newcommand{\RegFour}{$-$2.1}
\newcommand{\RegFourP}{$p<.001$}
\newcommand{\RegFourSE}{0.6}
\newcommand{\RegOne}{$-$3.5}
\newcommand{\RegThree}{$-$2.2}
\newcommand{\RegThreeAbs}{2.2}
\newcommand{\RegThreeP}{$p<.001$}
\newcommand{\RegThreeSE}{0.5}
\newcommand{\RegTwo}{$-$2.5}
\newcommand{\ResExtBinary}{91\%}
\newcommand{\ResSelfBinary}{92\%}
\newcommand{\RobustAllSig}{all}
\newcommand{\RobustMax}{$-$5.7}
\newcommand{\RobustMaxP}{$p=.033$}
\newcommand{\RobustMin}{$-$2.2}
\newcommand{\RobustN}{25}
\newcommand{\SameGapFinal}{$-$1.3}
\newcommand{\SameGapFinalSE}{0.3}
\newcommand{\SelfBrier}{0.105}
\newcommand{\SelfCalInt}{$-$0.38}
\newcommand{\SelfCalIntSE}{0.04}
\newcommand{\SelfDefaultOpen}{95\%}
\newcommand{\SelfEngagedGap}{$-$3.6}
\newcommand{\SelfGapFinal}{$-$4.4}
\newcommand{\SelfGapFinalAbs}{4.4}
\newcommand{\SelfGapFinalSE}{0.5}
\newcommand{\SelfGapOpen}{$-$4.8}
\newcommand{\SelfMeanPrice}{49.4\%}
\newcommand{\SelfRangeMidGap}{$-$9.4}
\newcommand{\SelfRangeMidGapAbs}{9.4}
\newcommand{\SelfRangeMidGapSE}{1.0}
\newcommand{\SelfReliability}{0.004}
\newcommand{\SelfResolution}{0.146}
\newcommand{\SelfSilentGap}{$-$7.4}
\newcommand{\SelfYesAtFifty}{39\%}
\newcommand{\SelfYesRate}{45.0\%}
\newcommand{\SilenceGapRegSelf}{$-$1.0}
\newcommand{\SilenceGapRegSelfSE}{1.2}
\newcommand{\StrangerShareBuys}{67\%}
\newcommand{\TrackB}{$-$0.60}
\newcommand{\TrackChronDiff}{$+$0.0007}
\newcommand{\TrackChronHi}{$+$0.0025}
\newcommand{\TrackChronLo}{$-$0.0012}
\newcommand{\TrackCreators}{286}
\newcommand{\TrackDiff}{$+$0.0001}
\newcommand{\TrackHi}{$+$0.0008}
\newcommand{\TrackHighPrice}{54\%}
\newcommand{\TrackHolm}{$p=.592$}
\newcommand{\TrackLo}{$-$0.0007}
\newcommand{\TrackLowPrice}{41\%}
\newcommand{\TrackN}{2,783}
\newcommand{\TrackP}{$p=.118$}
\newcommand{\TrackSE}{0.39}
\newcommand{\UnwantedDayPrice}{35\%}
\newcommand{\UnwantedRegAbs}{3.7}
\newcommand{\UnwantedRegHolm}{$p=.149$}
\newcommand{\UnwantedRegP}{$p=.021$}
\newcommand{\UnwantedRegSE}{1.6}
\newcommand{\ValAgreeControl}{65\%}
\newcommand{\ValCongRefers}{62\%}
\newcommand{\ValDesNeutralGap}{$-$5.6}
\newcommand{\ValDesNeutralGapSE}{4.1}
\newcommand{\ValDesUnwVsWantedAbs}{7.8}
\newcommand{\ValDesUnwVsWantedP}{$p=.111$}
\newcommand{\ValDesUnwVsWantedSE}{4.9}
\newcommand{\ValDesUnwantedGap}{$+$4.2}
\newcommand{\ValDesUnwantedGapHi}{$+$10.8}
\newcommand{\ValDesUnwantedGapLo}{$-$2.3}
\newcommand{\ValDesWantedGap}{$-$2.5}
\newcommand{\ValDesWantedGapSE}{2.8}
\newcommand{\ValEncPrecMax}{89\%}
\newcommand{\ValEncPrecMin}{85\%}
\newcommand{\ValEncRecallMax}{45\%}
\newcommand{\ValEncRecallMin}{43\%}
\newcommand{\ValEncRefers}{0.8\%}
\newcommand{\ValExtNotAboutMax}{99.0\%}
\newcommand{\ValExtNotAboutMin}{98.0\%}
\newcommand{\ValHcGap}{$-$1.9}
\newcommand{\ValHcGapSE}{1.6}
\newcommand{\ValHcN}{251}
\newcommand{\ValKappaAbout}{0.98}
\newcommand{\ValKappaCong}{0.95}
\newcommand{\ValKappaControl}{0.49}
\newcommand{\ValKappaDesir}{0.91}
\newcommand{\ValKappaEnc}{0.95}
\newcommand{\ValKappaTime}{0.93}
\newcommand{\ValKappaVerif}{0.22}
\newcommand{\ValPlatCreatorDet}{34}
\newcommand{\ValPlatN}{40}
\newcommand{\ValPlatProbMax}{0.5}
\newcommand{\ValRandomGap}{$-$2.9}
\newcommand{\ValRecomputeMaxRelOne}{0.02\%}
\newcommand{\ValRecomputeN}{23}
\newcommand{\ValRecomputeSERel}{1.7\%}
\newcommand{\ValSelfForwardMax}{99.5\%}
\newcommand{\ValSelfForwardMin}{98.5\%}
\newcommand{\ValSetAsideMax}{20\%}
\newcommand{\ValSetAsideMin}{18\%}
\newcommand{\ValUnwPrecMax}{90\%}
\newcommand{\ValUnwPrecMin}{86\%}
\newcommand{\ValUnwRecall}{47\%}
\newcommand{\ValUnwRecallHi}{66\%}
\newcommand{\ValUnwRecallLo}{35\%}
\newcommand{\ValUnwUK}{9}
\newcommand{\WMDiffOneCreator}{$+$4.8}
\newcommand{\WMDiffOneCreatorP}{$p=.003$}
\newcommand{\WMDiffOneCreatorSE}{1.6}
\newcommand{\WMExtCreatorManaLastDay}{61\%}
\newcommand{\WMExtOneCreator}{$-$0.9}
\newcommand{\WMExtOneCreatorSE}{1.0}
\newcommand{\WMExtOneExp}{$+$1.9}
\newcommand{\WMExtOneExpSE}{0.7}
\newcommand{\WMSelfCreatorManaLastDay}{62\%}
\newcommand{\WMSelfEncouragerPurchases}{1,276}
\newcommand{\WMSelfFiveCreator}{$-$0.7}
\newcommand{\WMSelfFiveCreatorSE}{1.7}
\newcommand{\WMSelfMarketsEncouragerAndStranger}{241}
\newcommand{\WMSelfOneCreatorAbs}{4.0}
\newcommand{\WMSelfOneCreatorP}{$p=.001$}
\newcommand{\WMSelfOneCreatorSE}{1.2}
\newcommand{\WMSelfOneEnc}{$+$0.7}
\newcommand{\WMSelfOneEncSE}{4.2}
\newcommand{\WMSelfOneExp}{$+$0.2}
\newcommand{\WMSelfOneExpSE}{0.8}
\newcommand{\WMSelfOnePrior}{$+$0.4}
\newcommand{\WMSelfOnePriorHi}{$+$2.0}
\newcommand{\WMSelfOnePriorLo}{$-$1.2}
\newcommand{\WMSelfOnePriorSE}{0.8}
\newcommand{\WMSelfThreeCreator}{$+$1.6}
\newcommand{\WMSelfThreeCreatorSE}{1.4}
\newcommand{\WMSelfTwoCreator}{$+$4.5}
\newcommand{\WMSelfTwoCreatorSE}{1.7}
\newcommand{\WTNo}{$+$0.8}
\newcommand{\WTNoSE}{1.8}
\newcommand{\WTTraders}{8,611}
\newcommand{\WTYes}{$-$0.3}
\newcommand{\WTYesSE}{1.8}

%% file: sections/00_abstract.tex
\begin{abstract}
People increasingly share personal goals online and invite others to forecast them. Research on the planning fallacy finds that observers predict a person's behavior less optimistically than the person does, and prediction markets aggregate such judgments, harnessing the wisdom of crowds. We study \NSelf{} prediction markets on Manifold in which \NSelfCreators{} users asked about their own futures (``Will I \ldots?''), reconstruct every order and execution, and compare these self-markets with markets about external events. Self-markets resolved YES \SelfGapFinalAbs{} percentage points less often than their final prices implied, whereas external markets were nearly calibrated (\ExtGapFinalAbs{} points); the difference was \RegThreeAbs{} points at the same price level and \RegFiveAbs{} points within creators who made both kinds of market. Absent when markets opened, the overstatement emerged within a day of trading, even where the person never traded, and peaked in the week before close. In a market's first day, observers moved external-market prices down toward the base rate, but moved self-market prices \FdDiffUpAbs{} points further toward YES, except in markets naming an outcome the person would not want, which were not overpriced. The correction came late and largely from the person, whose orders supplied \DecSelfCreatorShare{} of the eventual movement toward the outcome in self-markets, against \DecExtCreatorShare{} in external markets. Because self-market prices mainly failed to separate successes from failures, a recalibration learned from earlier markets improved later forecasts by only \HoAwareSelfRel{}. Crowds believe in the futures people pose for themselves, and the information that would correct them lies with the person.
\end{abstract}

%% file: sections/01_introduction.tex
\section{Introduction}
\label{sec:intro}

People increasingly announce their goals online and invite others to follow along: fitness apps share workouts with friends \citep{munson2012exploring,hamari2015likes}, habit trackers post streaks, and productivity communities run public accountability threads. Some platforms go further and let others bet. On the prediction-market platform Manifold, users routinely open markets on questions such as ``Will I go to the gym three times this week?'', ``Will I finish my thesis draft by Friday?'', or ``Will I stay off social media in March?'', and other users trade on them with the platform's currency. These markets turn an old question in social psychology, how well other people can predict what a person will do, into prices that can be checked against what happened.

Research on the planning fallacy gives a clear prediction. People forecasting their own behavior are optimistic: they underestimate how long their tasks will take, overestimate their own follow-through, and build their forecasts around their plans rather than their past \citep{buehler1994planning,koehler2006intentions,buehler2010planning}. Observers forecasting the same behavior lean on base rates and past behavior and are less optimistic \citep{buehler1994planning,epley2000holier,helzer2012peer}; when predicting generosity or test performance, they are also more accurate \citep{epley2000holier,helzer2012peer}. Prediction markets should add to this advantage, because trading pools the judgments of many observers and rewards those who correct mispricing \citep{wolfers2004prediction,arrow2008promise}. Markets about one person's future are thin, however, and price distortions in thin markets can persist for weeks \citep{rasooly2025manipulable}. Whether a market of observers forecasts a person well is therefore an empirical question.

We answer it with the public record of Manifold. From the \NCensus{} markets created on the platform between \CensusFirstYear{} and \CensusLastYear{}, we study \NSelf{} resolved self-markets, in which \NSelfCreators{} people asked about their own futures, and compare them with external markets, which concern events outside anyone's own life. We rebuild each market's price path and every executed purchase from the platform's order records. This lets us compare prices with outcomes at fixed points in a market's life, attribute each price change to the person or to observers, and compare the purchases of the person, of observers who had interacted with the person before, and of strangers in the same market. Two research assistants checked the reconstruction against the live platform, recomputed our estimates independently, and labelled a stratified sample of questions and comments to validate our coding.

Crowds believe in people too much. Self-markets resolved YES \SelfGapFinalAbs{} percentage points (pp) less often than their final prices implied. External markets made by other creators, by contrast, were close to calibrated: their outcomes occurred about as often as their prices implied. The difference held at the same price level and within creators who made both kinds of market. It was not present at the opening, when self-markets were, if anything, underpriced relative to external markets; it arose with trading. Within a day of opening, self-markets resolved YES less often than external markets at the same price, the difference was largest a week before close, and it appeared in markets in which the person never traded. In a market's first day, observers sold external markets down toward the lower rate at which such events occur, but they barely moved markets about people's own goals, and where the person never traded they did not move those prices down at all. The information that corrected prices came late and largely from the person, whose orders supplied \DecSelfCreatorShare{} of the movement toward the outcome in self-markets, against \DecExtCreatorShare{} in external markets. Markets naming outcomes the person would not want, such as relapsing or falling ill, were not overpriced, and observers who had interacted with the person before did not pick winning sides more often than strangers.

The study makes three contributions. First, it provides field evidence on how crowds forecast a specific person. The observer advantage documented in the lab did not appear in markets that pay for accuracy: prices overstated the futures people posed for themselves, the overstatement held across several benchmarks and at fixed forecasting times, and it emerged from observers' trading rather than from the person's opening. Second, it traces where the bias arises and what corrects it. Observers brought to events, but not to people, the outside view: the habit of judging a case by how similar cases usually turn out. The person, who holds the information that decides the outcome, supplied much of the correction near the end. The bias was absent where the question named an outcome the person would not want, and it did not depend on whether observers had interacted with the person before. Third, it offers practical and methodological guidance. Self-markets were less accurate mainly because their prices separated successes from failures less well, so a statistical correction recovers little; the missing ingredient is information held by the person. For research on trading platforms, we show that order records must be read as executions: treating each order as a trade distorts both prices and participation.

The paper proceeds as follows. Section~\ref{sec:background} reviews research on predicting one's own and other people's behavior and on prediction markets, and derives the research questions. Section~\ref{sec:method} describes the setting, the data, the reconstruction of prices and executions, and the measures. Section~\ref{sec:results} reports the results in the order of the research questions, and Section~\ref{sec:discussion} discusses their implications and limitations.

%% file: sections/02_background.tex
\section{Background and Research Questions}
\label{sec:background}

\subsection{Predicting one's own behavior and predicting someone else's}

People are poor forecasters of their own conduct. They underestimate how long their tasks will take \citep{buehler1994planning}, overestimate how often they will act generously \citep{epley2000holier}, and expect to follow through on intentions they later abandon \citep{koehler2006intentions,koehler2011intentions}. The planning-fallacy literature traces this optimism to how self-predictions are built: people construct a plan-based scenario for the task at hand and set aside relevant past experience, even when that experience shows similar plans failing \citep{buehler1994planning,newbyclark2000scenarios,buehler2010planning}. Current intentions receive heavy weight, while the factors that decide whether intentions become behavior receive little \citep{koehler2006intentions,sheeran2016gap}. Because intentions explain only part of later behavior \citep{sheeran2002intention}, a forecaster who reads intentions as outcomes will be too optimistic.

Observers are the classic remedy. In the original planning-fallacy studies, observers given information about another person's task predicted later completion times than the person did and drew more on past experience \citep{buehler1994planning}. Peers predicting someone's exam score or lab performance are less optimistic, and therefore more accurate, than the person, because they weight past achievement over aspiration \citep{helzer2012peer}. Predictions about other people's generosity are close to the truth while predictions about oneself are not, partly because people consult base rates when predicting others \citep{epley2000holier}. In the terms of \citet{kahneman1993timid}, observers adopt the outside view: they treat the case as one instance of a class and ask how such cases usually turn out, the logic that reference-class forecasting builds on \citep{flyvbjerg2008curbing}.

\subsection{Two ways observers can fall short}

The observer advantage rests on two conditions. The observer must be indifferent to the outcome, and the observer must have a reference class and a record to reason from. Each condition suggests a way in which observers could lose their advantage.

The first is motivation. Wanting an outcome can raise its judged probability \citep{weinstein1980unrealistic,kunda1990motivated}. Direct evidence for this desirability bias is more mixed than its familiarity suggests \citep{krizan2007desirability}, but it is clear in several settings. Incentives tied to quick completion make time predictions more optimistic by drawing attention to plans and away from past experience \citep{buehler1997motivated}, and desirability-driven optimism survives months of feedback and accuracy incentives \citep{massey2011hope,simmons2012optimism}. Bettors back the outcomes they want against the odds \citep{babad1991wishful}. Observers who know and like a person, or who have just read the person's hopeful plan, may forecast on that person's behalf as the person would. On this motivated account, the bias follows what the person wants: outcomes the person hopes for are overpriced, outcomes the person fears are underpriced, and the bias is strongest among the person's friends and supporters.

The second is a lack of information. Knowledge about a person is unevenly distributed: the person knows their own intentions and current progress, people close to them know their habits, and strangers know little of either \citep{vazire2010soka,vazire2008knowing}. A stranger asked whether this person will run every day this week knows little of the person's habits and has no obvious reference class. What the stranger does have is the question, which states the person's goal, and the person's own trades, if any. A forecaster who sees the plan but not the history behind it is in the position that, in the lab, makes actors optimistic. Describing a possibility explicitly also raises its judged probability \citep{tversky1994support}. On this informational account, observers' trades carry little information about the outcome, prices move toward the truth mainly when the person reveals what they know, and observers with more contact with the person forecast better than strangers.

\subsection{Prediction markets and the limits of aggregation}

Prediction markets are designed to correct individual errors. Traders who disagree with the price can profit by moving it, so prices aggregate dispersed information and tend to be well calibrated \citep{wolfers2004prediction,arrow2008promise}. Aggregation does not require every trader to be unbiased: in election markets many traders let their preferences color their trades, yet prices stay accurate because marginal traders set them \citep{forsythe1999wishes}, and traders who try to push prices away from the truth can even improve accuracy by attracting informed traders who profit from correcting them \citep{hanson2009manipulator}. Play-money markets forecast sports about as well as real-money markets \citep{servan2004money}, although real money improved accuracy for other events \citep{rosenbloom2006play}. Manifold's own play-money markets are well calibrated, with accuracy comparable to that of traditional prediction-market platforms \citep{rasooly2025manipulable}. In a large forecasting tournament, market prices were also more accurate than the simple average of forecasters' probability judgments \citep{atanasov2017distilling}.

Correction, however, requires informed traders who are willing to trade, and even active markets show systematic biases. Long shots are overpriced and favorites underpriced \citep{thaler1988parimutuel,snowberg2010favorite}, more so the longer the time to expiration \citep{page2013calibrated}. Prices can underreact to information when traders hold different beliefs and limited wealth \citep{ottaviani2015price}, and social influence can erode the independence that crowd accuracy relies on \citep{lorenz2011social}. On Manifold itself, randomly placed trades moved prices for weeks, and the effect was larger in markets with fewer traders and less volume \citep{rasooly2025manipulable}. A distortion that few informed traders oppose can therefore survive aggregation.

\subsection{Betting on a person}

Online platforms now let people turn their goals into public, priced forecasts. Commitment devices that put money or reputation at stake can change behavior \citep{bryan2010commitment,gine2010butt}, and personal-informatics and self-improvement systems use sharing, social feedback, and game elements to sustain effort \citep{munson2012exploring,hamari2015likes,sailer2017gamification,koivisto2019motivational}. Behavior spreads through online ties \citep{centola2010spread}, and making intentions public is double-edged: intentions that others notice can be enacted less intensively, because acknowledgment produces a premature sense of having reached the goal \citep{gollwitzer2009public}.

A market about one person's goal is an unusual forecasting object. Its subject controls the outcome, usually reports it, and can trade alongside everyone else. Its observers range from people who have followed the person for years to strangers who arrived through the platform's feed. Its question is the person's own framing of their future. The person's record on earlier goals is visible on their profile, but their progress on the current one is visible only through what they choose to say or trade. Such markets are also thin, which by the evidence above leaves room for prices to drift from the truth.

\subsection{Research questions}

Lab research predicts that a crowd of observers should be close to calibrated about a person's future, and if anything pessimistic. Both accounts above predict that it will not be, for different reasons, and they disagree about where the error lies and who could correct it. Four research questions follow.

The first question is whether the crowd overstates a person's future at all. RQ1 asks whether prices in markets about a person's own future overstate how often the named outcome occurs, relative to prices in markets about external events. If they do, the overstatement could be inherited from the opening price that the person sets, or it could arise as observers trade. RQ2 therefore asks when it arises (at the opening, with trading after the opening, or near the close) and whether it arises in markets in which the person never trades. The two accounts also disagree about who holds the information that would correct the price. RQ3 asks whose orders move self-market prices toward the eventual outcome, and how the person's purchases compare with those of observers in the same market. Finally, RQ4 confronts the accounts with the markers each emphasizes. It asks whether the overstatement is associated with the desirability of the named outcome and with the observers' relationship to the person, which the motivated account stresses; with cues that anyone can see, such as the person's earlier record and their silence in the market; with the person's history on the platform; or with the type of goal. Because a systematic overstatement could in principle be removed by a correction learned from past markets, we close by asking how much such a correction recovers. Of these questions, RQ1 and RQ2 are primary, RQ3 is secondary, and RQ4 is exploratory (Section~\ref{sec:analysis}).

%% file: sections/03_method.tex
\section{Method}
\label{sec:method}

\subsection{Setting}

Manifold is an online prediction-market platform on which any user can open a market on a yes-or-no question and other users trade on it with mana, the platform's currency. An automated market maker sets prices, which are read as the probability that the question resolves YES; traders can also post limit orders that rest until another order fills them. The market's creator writes the question, chooses the opening probability and closing date, may trade in the market like anyone else, and resolves it when the outcome is known. Markets are public, and every order is time-stamped and available through the platform's public interface \citep{manifold2026api}. Questions from the platform have been used as a forecasting benchmark \citep{halawi2024forecasting}, and its markets have been used for field experiments on price manipulation \citep{rasooly2025manipulable}.

Many users open markets about themselves: whether they will go to the gym three times this week, finish a draft by Friday, stay off social media for a month, or be admitted to a program. We call these self-markets, and we refer to the creator of a self-market, whose behavior is being forecast, as the person. The person controls the outcome, usually reports it, and can watch and join the trading. Everyone else in the market is an observer, forecasting another person's behavior with mana at stake.

\subsection{Data and sample}

We retrieved the full list of markets available through the platform's public interface on 22 and 23 October 2026 (\NCensus{} markets created from \CensusFirstYear{} to \CensusLastYear{}), and for the sampled markets every order and comment. All requests succeeded, and markets were deduplicated by their identifiers. A market is eligible when it is binary, denominated in mana, resolved YES or NO, had at least two unique traders according to the platform, and received at least one purchase by someone other than its creator before it closed. Markets resolved to a probability or cancelled have no binary outcome and are excluded; Appendix~\ref{app:sample} reports how often each kind of market ended this way and how the comparison would change if they were included.

A market is a self-market when its question asks, in the first person, about the creator's future: it begins with ``Will I'', ``Can I'', ``Shall I'', or ``Am I going to'', optionally after a bracketed tag or a short time phrase such as ``By Friday,''. The sampling frame used a broader first-person rule. Questions that look back (``Did I''), ask for advice (``Should I''), or ask about a present state (``Do I'', ``Am I'') were set aside (\NSelfDropped{} markets); including them leaves the results unchanged. External markets contain no first-person reference of any kind (no I, me, my, we, us, or our) and therefore concern events outside the creator's own life, such as elections, sports, technology releases, or platform milestones. Questions with other first-person content, such as markets about a creator's relatives or about ``we'', belong to neither group.

The sample contains every eligible self-market and two sets of external markets. The main benchmark is a random draw of external markets from creators who never opened a first-person market about themselves, stratified by creation quarter so that its timing matches that of the self-markets. The second set comprises up to \ExtCap{} randomly chosen external markets from every creator of a self-market. It shows how markets made by the same people behave when the question is not about them, and it supports comparisons within creator. Table~\ref{tab:sample} describes the \NSelf{} self-markets from \NSelfCreators{} creators, the \NExt{} external markets from \NExtCreators{} other creators, and the \NSame{} external markets from \NSameCreators{} self-market creators. Self-markets had fewer observers and less trading volume than external markets and more often ended at an intermediate price.

\input{generated/tab_sample}

\textbf{Research ethics.} All data are publicly visible on the platform and were collected through its documented public interface within its rate limits. Users appear under pseudonyms; we analyze aggregates only, report no usernames, and paraphrase rather than quote individual questions. The study involved no interaction with users. Because many questions concern health, relationships, and habits, the released data replace user identifiers with keyed hashes and contain no question or comment text; they include the market identifiers needed to rebuild the data from the platform.

\subsection{Reconstructing prices and executions}

The platform records orders, not trades. An order's record states the probability immediately before and after the order's own execution against the market maker, and its fills list what it bought or sold, when, and at what cost. A limit order may be filled long after it was placed, in several pieces, or partly filled and then cancelled; when it is filled later, the price change belongs to the incoming order that filled it. We therefore build two separate series. The price path is the sequence of all orders, including cancelled and unfilled ones, in the order they were placed. Executions are the fills, each at its own time, so that a cancelled order keeps what it had bought before cancellation and a resting order's purchase is dated when it was filled.

Two internal checks support this reconstruction (Appendix Table~\ref{tab:event}). The final price of the reconstructed path equals the probability the platform reports for the market in \EvFinalMatch{} of sample markets, against \EvFinalMatchOld{} when only filled, never-cancelled orders are used, and the probability before an order equals the probability after the preceding order in \EvContinuity{} of consecutive pairs, against \EvContinuityOld{}. The distinction matters for participation as well as prices: \EvPurchasesCancelledShare{} of all purchase executions came from orders that were later cancelled, and \EvLateFills{} purchase orders received a fill more than a minute after they were placed.

Two research assistants then checked the reconstruction externally. On 40 randomly drawn sample markets, the question text, the resolution, and the number of traders shown on the live platform matched our data in all \ValPlatN{} markets; the last probability on each market's price chart was within \ValPlatProbMax{} points of the reconstructed final price; and whether the creator had traded matched in all \ValPlatCreatorDet{} markets where the page showed it. Working only from the released de-identified data and with their own code, the assistants also recomputed \ValRecomputeN{} reported quantities, including the main regression coefficients and the purchase-level estimates. They reproduced every one to within \ValRecomputeMaxRelOne{}; one standard error differed by \ValRecomputeSERel{} because it was computed under another small-sample convention. Synthetic checks of partial fills, late fills, cancelled orders, and the estimators, together with checks of the processed data against the timing rules below, are released with the code.

\subsection{Measures}
\label{sec:measures}

The analyses rest on four kinds of measures: forecasts and their accuracy, the activity of the person and of observers, the roles of buyers, and the content of questions and comments.

\textbf{Outcomes and forecasts.} The outcome $Y$ equals 1 if the market resolved YES and 0 if it resolved NO. Trading in a market ends at the earlier of its scheduled close and its resolution, which we call its close. The main forecast is the final price $p$, the market probability after the last order before the close. Because the final price can follow news about the outcome, and because a market's life depends on when it resolved, we also measure prices at fixed points, or landmarks: at the opening; 1 hour, 6 hours, 1 day, 3 days, and 7 days after creation, for markets still open at that time; and 7 days and 1 day before the close, for markets that had opened by then. The price at a landmark is the probability after the last order at or before it. Three summaries of the price over the market's life complete the set: the price at its midpoint, the time-weighted average price, and the price after the third observer purchase.

\textbf{Calibration and accuracy.} For each market, the calibration gap is $Y-p$. Its mean is the difference between how often markets resolve YES and how often their prices say they will. A negative mean gap means that prices overstated YES, and we then call the prices optimistic. Because a mean gap can hide offsetting errors, we also report the Brier score, the mean squared difference between $Y$ and $p$ \citep{brier1950verification}; log loss; the intercept and slope of a logistic recalibration of $Y$ on the log-odds of $p$; and the Murphy decomposition of the Brier score into reliability (miscalibration), resolution (how well prices separate YES from NO outcomes), and uncertainty (the base rate) \citep{murphy1973partition}.

\textbf{Price changes and creator activity.} Each order's own price change, $\Delta p$, is attributed to the party that placed it: the creator, a human observer, or an automated observer, meaning an order placed through the platform's programming interface. Multiplying $\Delta p$ by $+1$ for markets that resolved YES and by $-1$ for markets that resolved NO measures movement toward the outcome; summed over a market's orders, it equals the reduction in the market's absolute error from the opening to the final price. Summing $\Delta p$ without the sign of the outcome measures movement toward YES. A creator is coded as never trading in a market when none of their orders placed up to the order that set the final price was executed, and as silent when they also posted no comment by that time. Both codes therefore use only activity that observers could see when the final price was set. The creator's position is the number of YES shares minus NO shares they held at that time. For each self-market, the creator's earlier record is the share of their earlier forward-looking self-markets, resolved before the focal market opened, that resolved YES, together with the number of such markets; because YES can name an unwanted outcome, we call this share the prior YES rate rather than a success rate. The record is visible to anyone who opens the creator's profile.

\textbf{Buyers and purchases.} Buyer roles are assigned at the time of each purchase and are mutually exclusive, in the following order. The creator is the market's creator. An encourager had posted a comment in the market, before placing the purchase order, that wished the creator well (for example ``good luck'', ``you got this'', ``rooting for you'', or similar emoji); congratulations, which can respond to a known outcome, are not counted. A prior interactor had traded or commented in another sampled market of the same creator before the focal market opened. Because only sampled markets are observed, prior interaction measures observed contact on the platform, not friendship. All remaining observers are strangers. Two further indicators describe observers: experienced observers had traded in 50 or more sampled markets before the focal market opened, and automated purchases were placed through the programming interface. For each purchase we record whether the purchased side won, and its payoff per share if held to resolution (1 if the purchased side won, 0 otherwise) minus the price paid per share. The payoff accounts for the price paid. Both measures describe purchases rather than beliefs, since a purchase also reflects timing, strategy, and the buyer's other positions.

\textbf{Question content.} Two keyword rules code self-market questions. The first flags questions whose YES answer is a lapse or an outcome the creator would not want, such as relapsing, breaking a streak, missing a deadline, falling ill, or losing money. It excludes negations, cessation goals (``Will I quit vaping?''), mixed outcomes (``promote or demote''), tests that come back negative, and questions about the creator's beliefs. The second separates behaviors the creator performs (effort goals such as exercising, writing, or finishing a task, and restraint goals such as quitting a habit) from outcomes that others or chance help decide (admissions, job offers, test scores, illness, winning). Both rules were refined by reading every flagged question and keyword-in-context listings of unflagged ones, fixed before validation, and released (Appendix~\ref{app:coding}).

\textbf{Validation of the coding.} Two research assistants labelled a stratified sample of 726 questions and 400 comments independently, without seeing market outcomes or the rules' labels (Appendix Table~\ref{tab:validation}). Agreement, measured by Cohen's $\kappa$ \citep{cohen1960coefficient}, was high on whether a question concerns the creator ($\kappa=\ValKappaAbout{}$), whether it concerns the future ($\kappa=\ValKappaTime{}$), and whether the creator would want a YES ($\kappa=\ValKappaDesir{}$), and on whether a comment encourages ($\kappa=\ValKappaEnc{}$) or congratulates the creator ($\kappa=\ValKappaCong{}$). The market classification held up: the coders labelled \ValSelfForwardMin{} to \ValSelfForwardMax{} of 200 random self-markets as forward-looking questions about the creator and \ValExtNotAboutMin{} to \ValExtNotAboutMax{} of 100 random external markets as not about the creator, whereas only \ValSetAsideMin{} to \ValSetAsideMax{} of the first-person questions set aside were forward-looking. The unwanted-outcome rule is precise but incomplete. Between \ValUnwPrecMin{} and \ValUnwPrecMax{} of the questions it flags were coded unwanted, and it flags an estimated \ValUnwRecall{} of the questions both coders labelled unwanted (95\% CI \ValUnwRecallLo{} to \ValUnwRecallHi{}). The encouragement rule is likewise precise (\ValEncPrecMin{} to \ValEncPrecMax{}) and incomplete (recall \ValEncRecallMin{} to \ValEncRecallMax{}). Only \ValEncRefers{} of the comments it flags showed knowledge of the outcome, against \ValCongRefers{} of those flagged as congratulating, which supports keeping congratulations out of the measure of encouragement. Goal type is the weakest measure: the coders agreed on how much an outcome depends on the creator for only \ValAgreeControl{} of questions ($\kappa=\ValKappaControl{}$). The analyses that use the keyword rules are therefore exploratory.

\subsection{Analysis}
\label{sec:analysis}

\textbf{Market-level comparisons.} Mean gaps and their differences are estimated by least squares with standard errors clustered by creator, because markets from the same creator share a person, a resolution style, and often a set of traders. Each market receives equal weight; a cap of \ExtCap{} markets per creator and equal weights per creator serve as robustness checks. The main comparison regresses the gap at the final price on a self-market indicator, with external markets by other creators as the benchmark. It then adds controls for the log number of observers, log purchase volume, log scheduled horizon, and creation-year fixed effects, and finally fixed effects for ten equal-width price bins (0--10\%, 10--20\%, \ldots, 90--100\%), which compare self- and external markets at the same price level. The controls describe the markets themselves, so the adjusted comparison is descriptive: it compares self- and external markets of similar size, horizon, timing, and price. A second set of specifications uses the same creators' external markets as the benchmark, with and without creator fixed effects. At each landmark we repeat the same-price comparison among the markets open at that time.

\textbf{Price changes and purchases.} Movement toward the outcome and toward YES is averaged over markets for each party. Purchase-level comparisons regress the indicator that the purchased side won on the role indicators, with strangers as the reference group and market fixed effects, so that each buyer is compared with strangers buying in the same market. Purchases are weighted by mana within each market, and each market receives equal total weight. The fixed effects are absorbed by subtracting weighted market means, which reproduces the estimates of a model with explicit market indicators; because markets are nested within creator clusters, the small-sample correction of the standard errors counts only the remaining regressors. Variants use market-by-day fixed effects, which compare purchases made on the same day of the same market; purchases made more than a day before the close; each buyer's first purchase in a market; equal weight per purchase; and the payoff per share as the outcome. Differences between self- and external markets are estimated jointly in a pooled model with interactions. Standard errors are clustered two ways, by creator and by buyer \citep{cameron2011robust}, because buyers trade across many creators' markets. A within-trader comparison, which compares the same observers' purchases in the two kinds of market at the same price paid, serves as a robustness check.

\textbf{Held-out evaluation.} To ask whether the bias can be corrected, we fit logistic recalibrations of the final price on markets created and resolved before 1 January 2025 and evaluate them on markets created afterwards, and separately in five folds that never share a creator between fitting and evaluation. Paired differences in Brier scores against the raw price come with intervals from a creator-cluster bootstrap.

\textbf{Inference.} The analyses were not preregistered. RQ1 and RQ2 are primary and RQ3 is secondary; the seven RQ4 tests are exploratory and are reported with $p$-values adjusted by Holm's method \citep{holm1979simple}. A difference that is not significant is reported with its confidence interval and is not taken as evidence of equivalence.

%% file: generated/tab_sample.tex
\begin{table}[t]
\caption{Analysis sample.}
\label{tab:sample}
\centering
\small
\begin{tabular}{lrrr}
\toprule
 & & \multicolumn{2}{c}{External markets} \\
\cmidrule(lr){3-4}
 & Self-markets & Other creators & Self-market creators \\
\midrule
Markets & 6,150 & 6,272 & 4,486 \\
Creators & 1,660 & 2,624 & 1,222 \\
Creation years & 2022--2026 & 2022--2026 & 2021--2026 \\
Share resolved YES & 45.0\% & 41.9\% & 40.0\% \\
Mean opening price & 49.8\% & 49.4\% & 49.7\% \\
Opened at the default 50\% & 95\% & 90\% & 95\% \\
Mean final price & 49.4\% & 42.8\% & 41.2\% \\
Final price between 20\% and 80\% & 37\% & 17\% & 18\% \\
Observers per market, median (mean) & 9 (13.6) & 12 (22.0) & 12 (25.3) \\
Observer purchase volume per market, median (mana) & 383 & 1,542 & 1,388 \\
Days from creation to close, median & 23.7 & 33.1 & 45.1 \\
Creator traded before the final price & 60\% & 60\% & 61\% \\
Creator commented before the final price & 50\% & 26\% & 38\% \\
Creator silent (no trade, no comment) & 22\% & 32\% & 25\% \\
At least one encouraging observer comment & 9.1\% & 0.5\% & 1.0\% \\
Observer purchases placed by programs & 20\% & 18\% & 17\% \\
Observer purchases filling a limit order & 25\% & 25\% & 23\% \\
\bottomrule
\end{tabular}
\par\smallskip
\begin{minipage}{0.9\textwidth}\footnotesize
Note: Binary, mana-denominated markets resolved YES or NO with at least one observer purchase before close. Self-markets ask a forward-looking question about the creator (``Will I \ldots?''); external markets contain no first-person reference. ``Other creators'' never opened a first-person market about themselves; ``self-market creators'' are the creators of the self-markets. Observers are traders other than the creator. Shares of purchases are averaged over markets.
\end{minipage}
\end{table}

%% file: sections/04_results.tex
\section{Results}
\label{sec:results}

\subsection{Prices overstate the futures people pose for themselves (RQ1)}
\label{sec:rq1}

\begin{figure}[t]
  \centering
  \includegraphics[width=\textwidth]{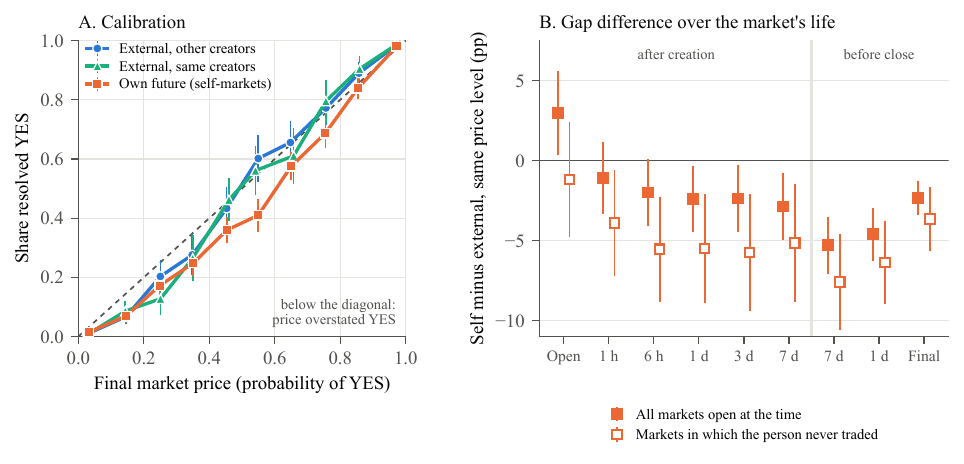}
  \caption{Calibration of self-markets and external markets. Panel A plots the share of markets that resolved YES against the final price, in ten equal-width price bins, with 95\% confidence intervals clustered by creator; points below the dashed diagonal mean that prices overstated YES. Panel B plots, at landmarks in the markets' lives, the difference between self-markets and external markets by other creators in the share resolving YES minus the price, at the same price level (fixed effects for ten price bins), with 95\% confidence intervals. Filled markers use all markets open at the time; hollow markers keep markets in which the person never traded before the final price.}
  \label{fig:calibration}
\end{figure}

Self-markets resolved YES in \SelfYesRate{} of cases, while their final prices averaged \SelfMeanPrice{}. The mean calibration gap is therefore \SelfGapFinal{} percentage points (pp; SE \SelfGapFinalSE{}). External markets made by other creators are close to calibrated, with a gap of \ExtGapFinal{} pp (SE \ExtGapFinalSE{}), so the difference between the two is \DiffGapFinal{} pp (SE \DiffGapFinalSE{}, \DiffGapFinalP{}). Figure~\ref{fig:calibration}A shows where the difference arises. Across the middle of the price range the self-market curve runs below the diagonal: according to the logistic recalibration, a self-market priced at one half resolved YES \SelfYesAtFifty{} of the time, against \ExtYesAtFifty{} for an external market by another creator. Near zero and one, both kinds of market are close to calibrated. Among markets whose final price lay between 20\% and 80\%, the gap is \SelfRangeMidGap{} pp for self-markets (SE \SelfRangeMidGapSE{}) and \ExtRangeMidGap{} pp for external markets by other creators (SE \ExtRangeMidGapSE{}); Appendix Table~\ref{tab:gaps} reports the gaps for other forecasts and price ranges.

Differences in the size, horizon, and timing of markets explain only part of the gap. Table~\ref{tab:regs} regresses the gap on a self-market indicator. Adding the number of observers, purchase volume, horizon, and creation-year fixed effects reduces the difference from \RegOne{} pp to \RegTwo{} pp, and adding price-bin fixed effects, which compare self- and external markets at the same price level, leaves \RegThree{} pp (SE \RegThreeSE{}, \RegThreeP{}). The gap is also not a matter of who makes the markets. External markets made by the people who also make self-markets behave like those of other creators, with a gap of \SameGapFinal{} pp (SE \SameGapFinalSE{}). Against these markets the self-market difference is \RegFour{} pp with controls and price-bin fixed effects (SE \RegFourSE{}, \RegFourP{}), and \RegFive{} pp within creator (SE \RegFiveSE{}, \RegFiveP{}), estimated from \RegFiveMarketsSelf{} self-markets by creators who made both kinds of market. The same people's markets were priced more optimistically when the question was about themselves than when it was about the world.

\input{generated/tab_regs}

The mean gap captures only one part of accuracy, and Table~\ref{tab:calib} reports the others. Self-markets' final prices had a Brier score of \SelfBrier{}, against \ExtBrier{} for external markets by other creators, and their calibration intercept was more negative (\SelfCalInt{}, SE \SelfCalIntSE{}, against \ExtCalInt{}, SE \ExtCalIntSE{}). The Murphy decomposition attributes most of the difference in Brier scores to resolution rather than reliability. Self-market prices separated the markets that resolved YES from those that resolved NO less well (resolution \SelfResolution{} against \ExtResolution{}), while their miscalibration, though larger, was small in absolute terms (reliability \SelfReliability{} against \ExtReliability{}). Self-market prices were thus both less informative and biased toward YES, and the bias is the systematic part of a larger shortfall in information.

\input{generated/tab_calib}

The answer to RQ1 is yes. At their final prices, markets about a person's own future overstated how often the named outcome occurred, by about \SelfGapFinalAbs{} pp on average and by about \SelfRangeMidGapAbs{} pp when the outcome was uncertain, whereas markets about external events were close to calibrated.

\subsection{The overstatement arises with trading (RQ2)}
\label{sec:rq2}

The final price is the last of many forecasts, and it can follow news about the outcome. Figure~\ref{fig:calibration}B and Appendix Table~\ref{tab:landmarks} therefore compare self- and external markets at the same price level at each landmark. At the opening, almost all markets of both kinds stood at the default 50\% (\SelfDefaultOpen{} of self-markets and \ExtDefaultOpen{} of external markets), and self-markets were, if anything, underpriced: at the same opening price they resolved YES \LmOpenDiffAbs{} pp more often than external markets (SE \LmOpenDiffSE{}). Trading reversed this. One hour after creation the difference was \LmHourDiff{} pp (SE \LmHourDiffSE{}), one day after creation \LmDayDiff{} pp (SE \LmDayDiffSE{}, \LmDayDiffP{}), and seven days after creation \LmWeekDiff{} pp (SE \LmWeekDiffSE{}, \LmWeekDiffP{}). It was largest in the final stretch: \LmWeekBeforeDiff{} pp a week before close (SE \LmWeekBeforeDiffSE{}) and \LmDayBeforeDiff{} pp a day before close (SE \LmDayBeforeDiffSE{}). A self-market priced at one half a week before close resolved YES \LmWeekBeforeSelfYesFifty{} of the time, against \LmWeekBeforeExtYesFifty{} for an external market at the same price.

The person's own trades are not needed for this pattern. In markets in which the person never traded before the final price, self-markets were priced like external markets at the opening (difference \LmOpenNeverDiff{} pp, SE \LmOpenNeverDiffSE{}). The overstatement appeared within an hour (\LmHourNeverDiff{} pp, SE \LmHourNeverDiffSE{}), reached \LmDayNeverDiff{} pp after a day (SE \LmDayNeverDiffSE{}), and stood at \LmWeekBeforeNeverDiff{} pp a week before close (SE \LmWeekBeforeNeverDiffSE{}). At the final price these markets were \NeverDiffAbs{} pp more optimistic than external markets in which the creator never traded (SE \NeverDiffSE{}, \NeverDiffP{}), a larger difference than for self-markets as a whole. Prices set by observers alone overstated the person's future from the first hours of trading.

The price changes in those first hours show how the overstatement arose. Table~\ref{tab:decomp} attributes each price change to the party whose order made it. In the first day after creation, human observers moved external-market prices \DayExtObsUpAbs{} pp toward NO on average, from the default 50\% toward the lower rate at which such questions resolve YES. In self-markets that did not name an unwanted outcome, they moved prices only \FdSelfWantedUpAbs{} pp toward NO (SE \FdSelfWantedUpSE{}), although these markets also resolved YES less than half the time (\FdSelfWantedYes{}). Observers' first-day movement toward YES was thus \FdDiffUpAbs{} pp larger in these self-markets than in external markets (SE \FdDiffUpSE{}, \FdDiffUpP{}), and \FdDiffUpCtrlAbs{} pp larger when the eventual outcome and the creation year are held fixed (SE \FdDiffUpCtrlSE{}). The contrast is sharpest where the person never traded. There, observers moved these self-markets \FdNeverSelfWantedUpAbs{} pp toward YES (SE \FdNeverSelfWantedUpSE{}), while they moved external markets with a similar YES rate (\FdNeverSelfWantedYes{} and \FdNeverExtYes{}) \FdNeverExtUpAbs{} pp toward NO, a difference of \FdNeverDiffUp{} pp (SE \FdNeverDiffUpSE{}, \FdNeverDiffUpP{}).

Observers did move one kind of self-market toward its base rate. In markets naming an unwanted outcome, they sold prices down by \FdSelfUnwantedUpAbs{} pp in the first day (SE \FdSelfUnwantedUpSE{}). Observers thus sold external events down toward their base rate, and sold down the misfortunes people named, but not the goals people set. Where the person did trade, the person's own first-day orders also moved prices toward YES (by \DaySelfCreatorUpAbs{} pp, against \DayExtCreatorUpAbs{} pp for the creators of external markets), and on balance these orders moved prices toward the eventual outcome (by \DaySelfCreatorTowardAbs{} pp).

\begin{figure}[t]
  \centering
  \includegraphics[width=\textwidth]{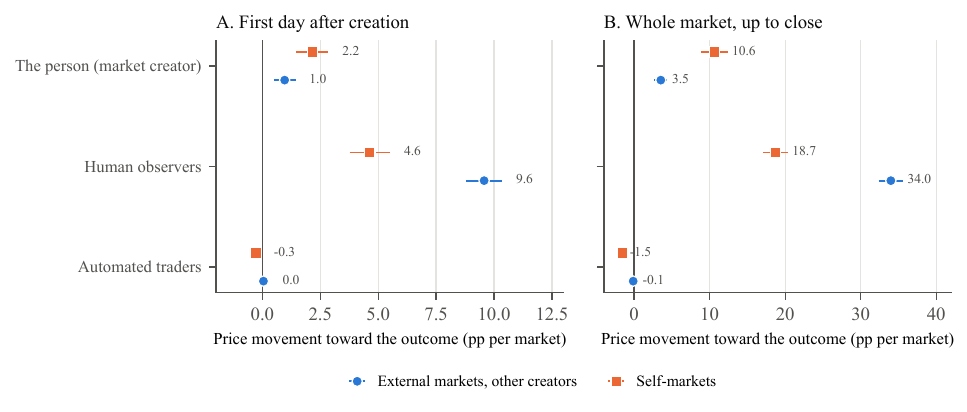}
  \caption{Whose orders moved prices toward the eventual outcome? Each order's own price change, signed so that positive values move the price toward the outcome, summed by the party that placed the order and averaged over markets, with 95\% confidence intervals clustered by creator. Panel A covers the first day after creation (markets open for at least a day); panel B covers the whole market up to close. External markets are those by other creators.}
  \label{fig:discovery}
\end{figure}

The answer to RQ2 is that the overstatement was not inherited from the opening price. It arose with trading within the first day, it arose without the person's trades, and it was largest in the week before close.

\subsection{The person supplies much of the correction (RQ3)}

If observers did not correct the overstatement early, the question is who did. Over a market's whole life, trading moved prices toward the outcome in both kinds of market, but less in self-markets: by \DecSelfTotalAbs{} pp on average, against \DecExtTotalAbs{} pp in external markets by other creators (Figure~\ref{fig:discovery}B, Table~\ref{tab:decomp}). The sources of that movement differed. Human observers' orders moved self-market prices \DecSelfObsAbs{} pp toward the outcome, against \DecExtObsAbs{} pp in external markets, and in the first day they moved self-market prices toward the outcome about half as far (\DaySelfObsTowardAbs{} against \DayExtObsTowardAbs{} pp). The person's own orders moved self-market prices \DecSelfCreatorAbs{} pp toward the outcome, against \DecExtCreatorAbs{} pp for the creators of external markets. The person thus supplied \DecSelfCreatorShare{} of the movement toward the outcome in self-markets, against \DecExtCreatorShare{} in external markets, and \DecSelfCreatorLateShare{} of the person's contribution came in the final day before close. Automated observers moved self-market prices slightly away from the outcome (\DecSelfAuto{} pp, SE \DecSelfAutoSE{}).

\input{generated/tab_decomp}

Purchases tell the same story. Table~\ref{tab:within} compares, within each market, how often each group's purchases landed on the side that won, relative to strangers buying in the same market. In self-markets, the person's purchases landed on the winning side \WMSelfOneCreatorAbs{} pp more often than strangers' (SE \WMSelfOneCreatorSE{}, \WMSelfOneCreatorP{}). In external markets, creators did no better than strangers (\WMExtOneCreator{} pp, SE \WMExtOneCreatorSE{}), and the difference between the two is \WMDiffOneCreator{} pp (SE \WMDiffOneCreatorSE{}, \WMDiffOneCreatorP{}). The advantage holds when purchases are compared within the same day of the same market (\WMSelfTwoCreator{} pp, SE \WMSelfTwoCreatorSE{}), but it is concentrated late: for purchases made more than a day before close it is \WMSelfThreeCreator{} pp (SE \WMSelfThreeCreatorSE{}), and it disappears once the price paid is taken into account (payoff per share \WMSelfFiveCreator{} pp, SE \WMSelfFiveCreatorSE{}). The person bought the side that won more often, mostly near the end and at prices that already reflected much of what they knew. Creators placed \WMSelfCreatorManaLastDay{} of their purchase volume in self-markets in the final day before close, much as creators of external markets did (\WMExtCreatorManaLastDay{}), but only in self-markets did these purchases land on the winning side more often than strangers'.

\input{generated/tab_within}

The person's advantage can come from knowing their own progress, from controlling the behavior, or from resolving the market. Whatever its source, the decisive information lay with the person and reached the price late. The person's position at the final price, which other traders can see, added little to the price in predicting the outcome (logistic coefficient for a YES position \CrePosYesB{}, SE \CrePosYesSE{}, \CrePosYesP{}).

\subsection{Desirability, prior interaction, and visible cues (RQ4, exploratory)}

The results so far show where the overstatement arises and who corrects it. RQ4 asks whether it follows the markers that the two accounts emphasize, beginning with desirability. The motivated account predicts that the bias follows what the person wants. Table~\ref{tab:accounts} splits self-markets by the named outcome. By the keyword rule, markets naming an outcome the person would want, or a neutral one, had a gap of \GoalWantedGap{} pp (SE \GoalWantedGapSE{}). The \GoalUnwantedN{} markets naming an unwanted outcome, such as relapsing, breaking a streak, or falling ill, had a gap of \GoalUnwantedGap{} pp (SE \GoalUnwantedGapSE{}; 95\% CI \GoalUnwantedGapLo{} to \GoalUnwantedGapHi{}); at the same price, they were \UnwantedRegAbs{} pp less optimistic than other self-markets (SE \UnwantedRegSE{}, \UnwantedRegP{}, Holm-adjusted \UnwantedRegHolm{}).

\input{generated/tab_accounts}

The coders' labels confirm that markets naming unwanted outcomes were not overpriced, and they leave open whether such markets were underpriced. The \ValHcN{} flagged markets that both coders confirmed as naming an unwanted outcome had a gap of \ValHcGap{} pp (SE \ValHcGapSE{}). Because the rule misses about half of unwanted questions, we also weighted the coded sample to the population of self-markets. These estimates rest mostly on the random draw of 200 self-markets, whose own gap was \ValRandomGap{} pp, against \SelfGapFinal{} pp for all self-markets. Markets the coders agreed named a wanted outcome had a gap of \ValDesWantedGap{} pp (SE \ValDesWantedGapSE{}), neutral or unclear ones \ValDesNeutralGap{} pp (SE \ValDesNeutralGapSE{}), and unwanted ones \ValDesUnwantedGap{} pp (95\% CI \ValDesUnwantedGapLo{} to \ValDesUnwantedGapHi{}). At the same price, unwanted outcomes were \ValDesUnwVsWantedAbs{} pp less optimistic than wanted ones (SE \ValDesUnwVsWantedSE{}, \ValDesUnwVsWantedP{}). A crowd siding with the person would produce this positive gap, but the estimate rests largely on \ValUnwUK{} unwanted questions that the rule missed in the random sample, each standing for many markets. These markets opened near 50\% although such outcomes occurred \GoalUnwantedYes{} of the time, and trading moved them down quickly, to \UnwantedDayPrice{} after a day. In these markets, siding with the person and applying the base rate both push prices down, so the two accounts make the same prediction.

Prior interaction did not distinguish observers. Observers who had interacted with the person on the platform before the market opened picked the winning side as often as strangers (\WMSelfOnePrior{} pp, SE \WMSelfOnePriorSE{}; 95\% CI \WMSelfOnePriorLo{} to \WMSelfOnePriorHi{}). Encouragement before a purchase was rare once comments made after the purchase were excluded (\WMSelfEncouragerPurchases{} purchases in the \WMSelfMarketsEncouragerAndStranger{} markets that also had strangers' purchases), and its estimate is too imprecise to interpret (\WMSelfOneEnc{} pp, SE \WMSelfOneEncSE{}). Most YES purchases in self-markets came from strangers (\StrangerShareBuys{} of observers' YES purchases). Experience on the platform helped with external markets (\WMExtOneExp{} pp, SE \WMExtOneExpSE{}) but not with self-markets (\WMSelfOneExp{} pp, SE \WMSelfOneExpSE{}).

The cues that anyone can see added little to the price. Prices followed a person's earlier record: self-markets by people in the lowest third of prior YES rates traded at \TrackLowPrice{} on average, against \TrackHighPrice{} for people in the highest third. Conditional on the price, however, the record added no detectable predictive value. In \TrackN{} self-markets by \TrackCreators{} creators with at least three earlier resolved self-markets, the coefficient on the prior YES rate in a logistic model with the log-odds of the price is \TrackB{} (SE \TrackSE{}, \TrackP{}, Holm-adjusted \TrackHolm{}). Adding the record changed out-of-sample Brier scores by \TrackDiff{} in creator-disjoint folds (95\% CI \TrackLo{} to \TrackHi{}) and by \TrackChronDiff{} when fitted on earlier and evaluated on later markets (\TrackChronLo{} to \TrackChronHi{}). Silence was priced in the same way. Silent self-markets, in which the person neither traded nor commented before the final price, had a larger raw gap (\SelfSilentGap{} pp against \SelfEngagedGap{} pp), but with price-bin fixed effects silence added \SilenceGapRegSelf{} pp (SE \SilenceGapRegSelfSE{}).

The person's history on the platform and the type of goal complete the picture. The gap did not shrink with a person's history, and if anything it was larger for people with long histories: among intermediate-priced self-markets it was \HistZeroMidGap{} pp for people with no earlier resolved self-market and \HistThirtyMidGap{} pp for people with thirty or more (log number of earlier self-markets, \HistReg{} pp, SE \HistRegSE{}). Because later markets attract different observers and different goals, this pattern describes heterogeneity across histories rather than a failure of the same observers to learn. Effort goals showed the largest gap, \GoalEffortGap{} pp (SE \GoalEffortGapSE{}), against \GoalOutcomeGap{} pp (SE \GoalOutcomeGapSE{}) for outcomes that others or chance help decide; with price-bin fixed effects, the difference between behavior goals and outcomes is \GoalReg{} pp (SE \GoalRegSE{}, \GoalRegP{}, Holm-adjusted \GoalRegHolm{}).

In sum, the exploratory tests found no overpricing in markets naming outcomes the person would not want, and no sign that prior interaction or visible cues account for the bias.

\subsection{How much can a correction help?}

Because the bias is systematic, a correction learned from past markets could in principle remove it. Table~\ref{tab:heldout} reports held-out Brier scores. Fitted on markets created and resolved before 2025 and evaluated on later ones, a recalibration that treats self-markets separately lowered the self-market Brier score by \HoAwareSelfRel{} (change \HoAwareSelfDiff{}, 95\% CI \HoAwareSelfLo{} to \HoAwareSelfHi{}); in creator-disjoint folds it lowered it by \HoCDAwareSelfRel{}. Applied to prices one day after creation, a pooled recalibration lowered the Brier score by \HoDayPooledSelfRel{} (95\% CI for the change \HoDayPooledSelfLo{} to \HoDayPooledSelfHi{}). The gains are small, and for final prices the chronological test cannot distinguish them from zero. The decomposition above explains why: most of what self-market prices lack is the ability to tell successes from failures, which no transformation of the price can add.

\input{generated/tab_heldout}

\subsection{Robustness}

Table~\ref{tab:robust} repeats the final-price comparison under \RobustN{} variations of the forecast, the sample, the benchmark, and the adjustment. The self-market difference ranges from \RobustMax{} to \RobustMin{} pp and is significant at the 5\% level in \RobustAllSig{} of them (largest $p$-value: \RobustMaxP{}). The forecast can be the price one day after creation or one day before close, the price at mid-life, the time-weighted average price, or the price after the third observer purchase. The sample can be restricted to markets whose final trade came within a day of close, to markets with at least seven or fifteen observers, to markets in which the creator never traded, or to either half of the sample period; it can exclude the ten most active self-market creators, markets in which a fifth or more of observer purchases were automated, same-day and same-week questions, and questions about the market itself or about chance devices such as coins and dice. Each creator can be limited to five markets of each kind or given equal weight. External questions that mention a user handle or a third-person pronoun can be dropped, and self-markets can be defined by the broad first-person rule. The benchmark can be the same creators' external markets or all external markets, and the adjustment can use twenty price bins or reweight external markets to resemble self-markets in size, horizon, year, opening, and price.

Two further checks address the selection of markets and of traders. Markets without a YES or NO resolution were about as common among self-markets as among external markets (\ResSelfBinary{} and \ResExtBinary{} resolved YES or NO; Appendix Table~\ref{tab:resol}), and counting them with a gap of zero leaves the self-market gap at \BndSelfAsPriced{} pp and the external gap at \BndExtAsPriced{} pp. A within-trader comparison does not detect a difference in the purchases of the same observers: among the \WTTraders{} observers who bought in both kinds of market, YES purchases in self-markets returned \WTYes{} pp per share relative to the same trader's YES purchases in external markets at the same price paid (SE \WTYesSE{}), and NO purchases \WTNo{} pp (SE \WTNoSE{}).

\input{generated/tab_robust}

%% file: generated/tab_regs.tex
\begin{table}[t]
\caption{Self-markets versus external markets: regressions of the calibration gap at the final price (percentage points).}
\label{tab:regs}
\centering
\small
\begin{tabular}{lccccc}
\toprule
 & (1) & (2) & (3) & (4) & (5) \\
\midrule
Self-market & $-$3.5 & $-$2.5 & $-$2.2 & $-$2.1 & $-$1.9 \\
 & (0.6) & (0.6) & (0.5) & (0.6) & (0.7) \\
\midrule
Comparison: external markets by & other & other & other & same & same \\
 & creators & creators & creators & creators & creators \\
Observers, volume, horizon (logs); year FE &  & Yes & Yes & Yes & Yes \\
Price-bin FE &  &  & Yes & Yes & Yes \\
Creator FE &  &  &  &  & Yes \\
Markets & 12,422 & 12,422 & 12,422 & 10,636 & 9,956 \\
Creators & 4,284 & 4,284 & 4,284 & 1,660 & 1,222 \\
\bottomrule
\end{tabular}
\par\smallskip
\begin{minipage}{0.92\textwidth}\footnotesize
Note: Dependent variable is $Y-p$ at the final price. Least squares with standard errors clustered by creator in parentheses. Columns (1)--(3) compare self-markets with external markets by creators who never opened a self-market; columns (4) and (5) compare them with external markets by the same creators, column (5) within creator (creators who made both kinds of market). The controls adjust the comparison descriptively (Section~\ref{sec:analysis}).
\end{minipage}
\end{table}

%% file: generated/tab_calib.tex
\begin{table}[t]
\caption{Accuracy and calibration of final prices.}
\label{tab:calib}
\centering
\small
\begin{tabular}{lrrr}
\toprule
 & & \multicolumn{2}{c}{External markets} \\
\cmidrule(lr){3-4}
 & Self-markets & Other creators & Same creators \\
\midrule
Brier score & 0.105 (0.005) & 0.049 (0.003) & 0.051 (0.002) \\
Log loss$^a$ & 0.326 (0.015) & 0.169 (0.009) & 0.176 (0.007) \\
Calibration intercept$^b$ & $-$0.38 (0.04) & $-$0.14 (0.05) & $-$0.18 (0.05) \\
Calibration slope$^c$ & 1.20 (0.04) & 1.32 (0.05) & 1.35 (0.05) \\
Implied YES rate at a price of 50\% & 39\% & 47\% & 46\% \\
Reliability (miscalibration)$^d$ & 0.004 & 0.001 & 0.001 \\
Resolution (discrimination)$^d$ & 0.146 & 0.195 & 0.190 \\
Uncertainty (base rate)$^d$ & 0.247 & 0.243 & 0.240 \\
\bottomrule
\end{tabular}
\par\smallskip
\begin{minipage}{0.9\textwidth}\footnotesize
Note: Lower Brier score and log loss are better. Standard errors clustered by creator in parentheses. $^a$Prices clipped to [0.01, 0.99]. $^b$Intercept of a logistic model of $Y$ with the log-odds of the price as an offset; negative values mean that prices overstated YES on average. $^c$Coefficient on the log-odds of the price; values above 1 mean that outcomes were more extreme than prices. $^d$Murphy decomposition over ten equal-width price bins: Brier score $\approx$ reliability $-$ resolution $+$ uncertainty.
\end{minipage}
\end{table}

%% file: generated/tab_decomp.tex
\begin{table}[t]
\caption{Whose orders moved prices toward the outcome? Price changes per market, in percentage points.}
\label{tab:decomp}
\centering
\footnotesize
\begin{tabular}{lrrr}
\toprule
 & & \multicolumn{2}{c}{External markets} \\
\cmidrule(lr){3-4}
 & Self-markets & Other creators & Same creators \\
\midrule
Total movement from opening to final price & $+$27.8 (0.9) & $+$37.3 (0.9) & $+$37.7 (0.4) \\
\quad by the creator's orders & $+$10.6 (0.9) & $+$3.5 (0.5) & $+$5.1 (0.5) \\
\quad by human observers' orders & $+$18.7 (0.8) & $+$34.0 (0.8) & $+$33.5 (0.6) \\
\quad by automated orders & $-$1.5 (0.2) & $-$0.1 (0.3) & $-$0.9 (0.3) \\
\quad not attributed to an order & $-$0.1 (0.0) & $-$0.1 (0.0) & $+$0.0 (0.0) \\
Creator, more than a day before close & $+$5.6 (0.5) & $+$1.7 (0.3) & $+$2.8 (0.4) \\
Creator, final day before close & $+$5.0 (0.7) & $+$1.8 (0.3) & $+$2.3 (0.2) \\
Observers, more than a day before close & $+$12.9 (0.7) & $+$23.0 (0.8) & $+$23.6 (0.6) \\
Observers, final day before close & $+$5.9 (0.3) & $+$11.0 (0.6) & $+$9.9 (0.4) \\
\midrule
\multicolumn{4}{l}{First day after creation (markets open for at least a day)} \\
\quad Creator, toward the outcome & $+$2.2 (0.4) & $+$1.0 (0.2) & $+$1.3 (0.2) \\
\quad Human observers, toward the outcome & $+$4.6 (0.4) & $+$9.6 (0.4) & $+$9.8 (0.4) \\
\quad Automated, toward the outcome & $-$0.3 (0.1) & $+$0.0 (0.1) & $+$0.1 (0.1) \\
\quad Creator, toward YES & $+$2.9 (0.4) & $+$1.0 (0.3) & $+$0.5 (0.3) \\
\quad Human observers, toward YES & $-$1.1 (0.4) & $-$4.5 (0.5) & $-$4.6 (0.4) \\
\quad Automated, toward YES & $+$0.9 (0.1) & $-$0.4 (0.1) & $-$0.1 (0.1) \\
\bottomrule
\end{tabular}
\par\smallskip
\begin{minipage}{0.92\textwidth}\footnotesize
Note: Each order's own price change, $\Delta p$, is multiplied by $+1$ if the market resolved YES and $-1$ if it resolved NO and summed by the group that placed the order; summed over all orders, this equals the reduction in the market's absolute error from the opening to the final price. ``Toward YES'' rows sum $\Delta p$ without the sign of the outcome. Means over markets, with standard errors clustered by creator in parentheses.
\end{minipage}
\end{table}

%% file: generated/tab_within.tex
\begin{table}[t]
\caption{Purchases on the side that won, relative to strangers buying in the same market (percentage points).}
\label{tab:within}
\centering
\footnotesize
\begin{tabular}{lrrr}
\toprule
 & Self-markets & External markets & Difference \\
\midrule
The creator & $+$4.0 (1.2) & $-$0.9 (1.0) & $+$4.8 (1.6) \\
Observers with a prior interaction & $+$0.4 (0.8) & $-$1.2 (0.8) & $+$1.6 (1.1) \\
Observers who had encouraged & $+$0.7 (4.2) & $+$6.0 (5.9) & $-$5.2 (7.4) \\
Experienced observers & $+$0.2 (0.8) & $+$1.9 (0.7) & $-$1.7 (0.9) \\
Automated purchases & $-$17.5 (2.1) & $-$16.7 (1.4) & $-$0.8 (2.5) \\
\midrule
Purchases & 188,201 & 707,742 & \\
Markets & 6,150 & 10,758 & \\
Markets with creator and stranger purchases & 3,499 & 6,297 & \\
\midrule
\multicolumn{4}{l}{The creator under alternative specifications} \\
\quad Market-by-day fixed effects & $+$4.5 (1.7) & $+$0.1 (1.4) & $+$4.4 (2.2) \\
\quad Purchases more than a day before close & $+$1.6 (1.4) & $-$3.0 (1.1) &  \\
\quad Each buyer's first purchase in the market & $+$3.1 (1.8) & $-$4.3 (1.4) &  \\
\quad Equal weight per purchase & $+$9.6 (2.5) & $-$5.2 (2.1) &  \\
\quad Payoff minus price paid, market-by-day fixed effects & $-$0.7 (1.7) & $+$0.8 (1.3) &  \\
\bottomrule
\end{tabular}
\par\smallskip
\begin{minipage}{0.92\textwidth}\footnotesize
Note: Weighted least squares of an indicator that the purchased side won on buyer-role indicators with market fixed effects, estimated by the weighted within transformation. Roles are mutually exclusive in the order creator, encourager (had posted an encouraging comment in the market before placing the purchase order), prior interactor (had traded or commented in another sampled market of the creator before the market opened); strangers are the reference. Indicators for experienced observers (50 or more earlier sampled markets) and automated purchases enter alongside. Purchases are weighted by mana within each market and each market receives equal total weight, except where noted. The difference is the self-market interaction in a pooled model. Standard errors two-way clustered by creator and buyer in parentheses. The last row uses the hold-to-resolution payoff per share minus the price paid.
\end{minipage}
\end{table}

%% file: generated/tab_accounts.tex
\begin{table}[t]
\caption{Self-markets by named outcome, goal type, and the creator's earlier record (exploratory).}
\label{tab:accounts}
\centering
\small
\begin{tabular}{lrrrr}
\toprule
 & Markets & Resolved YES & Mean final price & Gap (pp) \\
\midrule
Named outcome wanted or neutral & 5,856 & 45.9\% & 50.5\% & $-$4.6 (0.5) \\
Named outcome unwanted (lapse, illness, loss) & 294 & 27.2\% & 27.7\% & $-$0.5 (1.5) \\
\midrule
Effort goal (exercise, write, finish) & 2,236 & 46.2\% & 52.5\% & $-$6.3 (1.0) \\
Restraint goal (quit, stay off) & 71 & 52.1\% & 55.0\% & $-$2.9 (3.9) \\
Outcome others or chance help decide & 1,345 & 42.5\% & 45.8\% & $-$3.3 (0.9) \\
Unclassified & 2,498 & 45.1\% & 48.5\% & $-$3.4 (0.6) \\
\midrule
Prior YES rate, lowest third (28\%) & 928 & 37.6\% & 41.5\% & $-$3.9 (1.1) \\
Prior YES rate, middle third (50\%) & 927 & 41.6\% & 48.5\% & $-$6.9 (1.4) \\
Prior YES rate, highest third (75\%) & 928 & 50.1\% & 54.4\% & $-$4.3 (1.3) \\
\bottomrule
\end{tabular}
\par\smallskip
\begin{minipage}{0.92\textwidth}\footnotesize
Note: Gap is the share resolved YES minus the mean final price, with creator-clustered standard errors in parentheses. Named-outcome and goal codes come from released keyword rules applied to the question text. The last panel uses self-markets whose creator had at least three earlier forward-looking self-markets resolved before the market opened; the prior YES rate is the share of those that resolved YES (in parentheses, the mean within the third).
\end{minipage}
\end{table}

%% file: generated/tab_heldout.tex
\begin{table}[t]
\caption{Held-out Brier scores of market prices and simple corrections.}
\label{tab:heldout}
\centering
\footnotesize
\setlength{\tabcolsep}{4pt}
\begin{tabular}{lrrrr}
\toprule
 & \multicolumn{2}{c}{Self-markets} & \multicolumn{2}{c}{External markets, other creators} \\
\cmidrule(lr){2-3}\cmidrule(lr){4-5}
Forecast & Brier & Change [95\% CI] & Brier & Change [95\% CI] \\
\midrule
\multicolumn{5}{l}{Chronological split, final price} \\
\quad Market price & 0.141 &  & 0.062 &  \\
\quad Base rate by market type & 0.249 & 0.1079 [0.0906, 0.1259] & 0.246 & 0.1839 [0.1720, 0.1956] \\
\quad Pooled recalibration & 0.139 & $-$0.0015 [$-$0.0041, 0.0012] & 0.062 & $-$0.0002 [$-$0.0019, 0.0016] \\
\quad Self-aware recalibration & 0.139 & $-$0.0021 [$-$0.0058, 0.0022] & 0.061 & $-$0.0009 [$-$0.0020, 0.0001] \\
\quad Self-aware with market features & 0.139 & $-$0.0016 [$-$0.0058, 0.0032] & 0.061 & $-$0.0006 [$-$0.0019, 0.0006] \\
\multicolumn{5}{l}{Chronological split, price one day after creation} \\
\quad Market price & 0.230 &  & 0.183 &  \\
\quad Base rate by market type & 0.248 & 0.0181 [0.0061, 0.0310] & 0.243 & 0.0603 [0.0490, 0.0709] \\
\quad Pooled recalibration & 0.224 & $-$0.0057 [$-$0.0114, $-$0.0008] & 0.184 & 0.0012 [$-$0.0027, 0.0051] \\
\quad Self-aware recalibration & 0.224 & $-$0.0056 [$-$0.0122, 0.0008] & 0.183 & $-$0.0000 [$-$0.0032, 0.0029] \\
\quad Self-aware with market features & 0.226 & $-$0.0041 [$-$0.0117, 0.0030] & 0.183 & 0.0003 [$-$0.0028, 0.0034] \\
\multicolumn{5}{l}{Creator-disjoint folds, final price} \\
\quad Market price & 0.105 &  & 0.049 &  \\
\quad Base rate by market type & 0.248 & 0.1426 & 0.244 & 0.1950 \\
\quad Pooled recalibration & 0.102 & $-$0.0030 & 0.048 & $-$0.0004 \\
\quad Self-aware recalibration & 0.102 & $-$0.0037 & 0.048 & $-$0.0008 \\
\quad Self-aware with market features & 0.102 & $-$0.0035 & 0.048 & $-$0.0009 \\
\bottomrule
\end{tabular}
\par\smallskip
\begin{minipage}{0.95\textwidth}\footnotesize
Note: Chronological split: corrections are fitted on markets created and resolved before 1 January 2025 (11,435 markets) and evaluated on markets created afterwards (1,297 self-markets and 1,315 external markets by other creators). Creator-disjoint folds: five folds in which no creator appears in both training and evaluation; changes are pooled over folds. Recalibration models are logistic regressions on the log-odds of the price; the self-aware model lets the intercept and slope differ for self-markets; the feature model adds, for self-markets, whether the creator had traded and the log number of observers, both measured as of the forecast (by the final price, or within the first day). Changes are relative to the market price; negative changes are improvements. Intervals from a creator-cluster bootstrap of the paired score difference.
\end{minipage}
\end{table}

%% file: generated/tab_robust.tex
\begin{table}[t]
\caption{Robustness of the self-market difference in calibration gaps (percentage points).}
\label{tab:robust}
\centering
\footnotesize
\setlength{\tabcolsep}{4pt}
\begin{tabular}{llrrr}
\toprule
 & Variation & Difference & $N$ self & $N$ total \\
\midrule
Forecast & Baseline: final price, no controls & $-$3.5 (0.6) & 6,150 & 12,422 \\
 & Price at mid-life & $-$4.8 (0.9) & 6,150 & 12,422 \\
 & Time-weighted average price & $-$4.5 (0.8) & 6,150 & 12,422 \\
 & Price after third observer purchase & $-$2.2 (1.0) & 5,757 & 11,702 \\
 & Price one day after creation & $-$3.7 (1.0) & 5,383 & 10,820 \\
 & Price one day before close & $-$4.8 (0.8) & 5,383 & 10,820 \\
\midrule
Sample & Final trade within a day of close & $-$2.8 (0.6) & 4,308 & 8,850 \\
 & At least 7 observers & $-$3.2 (0.5) & 4,024 & 8,622 \\
 & At least 15 observers & $-$2.3 (0.7) & 1,673 & 4,228 \\
 & At most 5 markets per creator and type & $-$4.4 (0.7) & 3,480 & 7,838 \\
 & Equal weight per creator within type & $-$2.9 (0.8) & 6,150 & 12,422 \\
 & Without the 10 most active self-market creators & $-$3.4 (0.6) & 5,081 & 11,353 \\
 & Automated purchases under 20\% & $-$2.8 (0.7) & 3,429 & 7,236 \\
 & Creator never traded & $-$5.7 (1.0) & 2,466 & 5,004 \\
 & Excluding same-day and same-week questions & $-$3.5 (0.5) & 5,337 & 11,457 \\
 & Excluding self-referential and chance questions & $-$3.6 (0.6) & 5,982 & 12,254 \\
 & External questions without person mentions & $-$3.7 (0.6) & 6,150 & 12,196 \\
 & Created 2021--2023 & $-$3.1 (0.8) & 3,603 & 7,281 \\
 & Created 2024--2026 & $-$4.2 (0.9) & 2,547 & 5,141 \\
 & Broad self-market definition (all first-person leads) & $-$3.5 (0.6) & 6,270 & 12,542 \\
\midrule
Benchmark & Benchmark: same creators' external markets & $-$3.1 (0.6) & 6,150 & 10,636 \\
 & Benchmark: all external markets & $-$3.4 (0.6) & 6,150 & 16,908 \\
\midrule
Adjustment & Controls and 10 price-bin effects & $-$2.2 (0.5) & 6,150 & 12,422 \\
 & Controls and 20 price-bin effects & $-$2.2 (0.5) & 6,150 & 12,422 \\
 & Reweighted external markets (propensity, trimmed) & $-$3.5 (0.9) & 6,150 & 12,421 \\
\bottomrule
\end{tabular}
\par\smallskip
\begin{minipage}{0.92\textwidth}\footnotesize
Note: Coefficient on a self-market indicator in a regression of $Y-p$ (without controls unless stated); standard errors clustered by creator in parentheses. Unless stated otherwise, $p$ is the final price and the benchmark is external markets by other creators. The reweighting row weights external markets by the odds of a propensity score built from observers, volume, horizon, year, default opening, and price bin, after trimming scores outside [0.05, 0.95].
\end{minipage}
\end{table}

%% file: sections/05_discussion.tex
\section{Discussion}
\label{sec:discussion}

\subsection{Theoretical implications}

Lab studies of self- and social prediction cast observers as the corrective to optimistic self-prediction: free of the actor's plans and hopes, they consult base rates and past behavior \citep{buehler1994planning,epley2000holier,helzer2012peer}. In the markets studied here, observers aggregated through a market and paid for accuracy did not play that role. They brought an outside view to external events, selling prices down toward the rate at which such events occur, but not to the goals people set for themselves, and markets about people ended up overstating those goals. The observer advantage evidently depends on conditions that online crowds forecasting a specific person do not meet. Lab observers share a task with the actor, see the actor's history, and judge one person at a time. Observers on a platform see a question, a price, perhaps the person's own trades, and at most a profile. With a plan in view and little else about the person at hand, they are in the position that makes actors optimistic in the lab, and their prices look like the actor's forecast: the outcome the person aims for is treated as more likely than it is.

Why observers price people this way is the question our data cannot settle. Two accounts fit parts of the evidence. On a motivated account, observers share the person's hopes; consistent with it, the bias was absent from markets naming outcomes the person would not want, in the first day observers held up the prices of people's goals while selling down the prices of the misfortunes they named, and by the coders' labels unwanted outcomes were, if anything, underpriced. On an informational account, observers read a stated goal as evidence of what will happen and have nothing better to go on; consistent with it, the unwanted outcomes the rule flags and the coders confirmed were not underpriced, prior interactors traded like strangers, and prices moved toward the truth when the person, who holds the decisive information, traded. Both accounts agree that observers lacked, or did not use, a reference class for the person. Separating them requires holding information fixed while varying the observer's stake in the person, the kind of experiment described in Section~\ref{sec:limits}.

The findings also qualify the aggregation argument for prediction markets. Markets correct individual errors when informed traders oppose them \citep{forsythe1999wishes,wolfers2004prediction,hanson2009manipulator}. In self-markets the best-informed trader is a single person who may not trade, and who learns most about the outcome near the end. Most observer purchases came from strangers, and in a market's first day observers' trades moved self-market prices only about half as far toward the outcome as external-market prices, so the correction arrived late and incompletely. This matches the field evidence that price distortions on the same platform persist when few traders are present \citep{rasooly2025manipulable}, and it adds a distinct reason for a distortion to survive: the information that would remove it is held by the person whose future is being priced.

Finally, the accuracy of forecasts about people depends more on information than on bias. Self-markets were biased toward YES, but most of their shortfall in accuracy came from prices that separated successes from failures less well, and a statistical correction learned from past markets recovered little of it. The visible cues that might stand in for private information, such as a person's earlier record and their silence, added no detectable predictive value beyond the price. Forecasts about people may therefore be harder to improve by disclosing public information or by correcting bias than forecasts about events, because what they lack is held privately by the person.

\subsection{Practical implications}

Systems that make personal goals public, including commitment contracts, accountability communities, social fitness apps, and prediction platforms \citep{bryan2010commitment,munson2012exploring,hamari2015likes,koivisto2019motivational}, increasingly show users what others expect of them. Three implications follow, each a design hypothesis to be tested rather than a demonstrated remedy. First, crowd expectations about a person's goal can run optimistic, and designers who display them could show, alongside the crowd's view, how often similar goals by similar users succeeded. Second, because the person holds the information that decides the outcome, features that invite the person to report progress, or to state their own probability, may improve such forecasts more than features that attract more observers. Third, the bias is an average over markets, and a market's price remains informative about a person's chances even though prices run high on average. Publicly noticed intentions can feel like progress \citep{gollwitzer2009public}, and whether optimistic crowd signals help or hurt goal pursuit is an open question for such systems.

For prediction platforms, markets on individuals' own conduct differ from markets on public events in a way that matters for design. The person whose behavior is forecast trades with information others lack and usually decides the resolution. In our data, the person's purchases landed on the winning side more often than strangers' only in self-markets, mostly near the close. Rules about whether and when the subject may trade, how progress is disclosed, and how resolutions are verified therefore shape both the accuracy of the price and its fairness to other traders.

\subsection{Limitations and future research}
\label{sec:limits}

Several features of the setting open directions for further work. The first concerns how far the findings generalize. The markets use play money, and the platform's users are a particular online community. Play-money markets forecast sports about as well as real-money markets \citep{servan2004money}, but real money improved accuracy for other events \citep{rosenbloom2006play}, so replicating the analysis with real stakes, and with users of mainstream goal-tracking or social platforms, would show how far the pattern extends. The analysis is also limited to markets that resolved YES or NO. Markets that were cancelled or resolved to a probability were about as common among self-markets as among external markets, but their outcomes are unknown.

The second concerns measurement. The classification of questions and the measures of named outcomes, goals, and encouragement rest on keyword rules. Independent coding confirmed the market classification, but it showed that the unwanted-outcome and encouragement rules find only about half of the cases they target, and that a question's title says too little to code goal types reliably. Larger coded samples, or coding that draws on market descriptions, would sharpen the comparisons of desirability and goal type, which here rest on few unwanted questions and on a coarse goal measure. Prior interaction is measured within the sampled markets, so it captures observed contact on the platform rather than friendship, and encouragement before a purchase was too rare to evaluate; data that link platform activity to offline relationships would show whether people who know the person forecast better. Finally, the person usually reports the outcome, and late purchases can reflect progress, control over the behavior, or resolution practices. Whether others could verify an outcome could not be coded reliably from titles ($\kappa=\ValKappaVerif{}$), and studies of disputes and moderator corrections could assess how reporting affects measured calibration.

The third concerns mechanism. The design is observational, and it cannot hold observers' information fixed while their stake in the person varies. Experiments that elicit probabilities from the person and from observers at the same moment, vary how a question is framed (naming success or failure), display a reference class, or vary observers' access to the person's progress separately from their stake in the person could test the accounts discussed here and the remedies they suggest.

%% file: sections/06_conclusion.tex
\section{Conclusion}
\label{sec:conclusion}

When people ask the crowd about their own futures, the crowd believes them too readily. Across \NSelf{} prediction markets about people's own goals and prospects, prices overstated the futures people posed for themselves, except when the question named an outcome the person would not want, while markets about external events were close to calibrated. The overstatement was not present when markets opened; it arose as observers traded, appeared where the person never traded, and was largest in the week before close. Observers brought the outside view to events but not to people, and the information that corrected prices came late and largely from the person. Platforms that turn personal goals into public forecasts should not expect a crowd to supply the base rates that lab observers supply, and should look to the person for the information that the crowd does not have.

%% file: sections/A_appendix.tex
\section{Sample construction and reconstruction}
\label{app:sample}

Table~\ref{tab:flow} traces the sample from the full list of markets retrieved from the platform to the analysis sample. Table~\ref{tab:event} reports the checks of the reconstructed prices and executions described in Section~\ref{sec:method}. Table~\ref{tab:resol} compares how often self-markets and external markets ended without a YES or NO resolution, among binary, mana-denominated markets with at least two traders that closed before 1 July 2026, and shows the mean gap at each market's last price when such markets are added with a gap of zero, or when all of them are assigned NO or all YES.

\input{generated/tab_flow}

\input{generated/tab_event}

\input{generated/tab_resol}

\FloatBarrier

\section{Coding rules}
\label{app:coding}

\textbf{Market classification.} A question is a self-market when it opens, after an optional bracketed tag and an optional short time phrase followed by a comma (beginning with by, before, on, in, this, next, today, tonight, or tomorrow), with ``will I'', ``can I'', ``shall I'', or ``am I going to''. The sampling frame used a broader lead that also admitted would, could, do, does, am, should, and did; those questions are set aside in the analysis and used in a robustness check. A question is an external market when it contains none of the words I, me, my, mine, myself, I'm, I'll, I've, I'd, we, us, our, ours, or ourselves. All remaining questions, for example those about a creator's relatives or those phrased with ``we'', belong to neither group. External questions can still concern named people; dropping external questions that mention a user handle, the word ``creator'', or a third-person pronoun leaves the results unchanged (Table~\ref{tab:robust}).

\textbf{Unwanted named outcomes.} A self-market question is flagged when YES names a lapse or an adverse event for the creator: relapsing, breaking a streak, diet, or promise; losing a game, a job, money, mana, or a possession; failing; missing a class, a flight, or a deadline; falling sick or testing positive; injury, hospitalization, or an accident; being fired, laid off, banned, suspended, or arrested; regret; procrastinating, oversleeping, skipping, or staying up late; drinking alcohol, smoking, vaping, or bingeing; going over budget; breaking up or dropping out; and demotion. The flag is withdrawn when the question negates the term or asks about stopping or avoiding it (not, without, never, quit, stop, avoid), names a mixed outcome (``promote or demote'', ``lose or gain''), asks about a negative test, a vaccine, or a recovery, or asks about the creator's beliefs (believe, agree, convinced). The rules were refined by reading every question flagged by an earlier version and keyword-in-context listings of unflagged questions containing related terms.

\textbf{Goal content.} The goal-type rule assigns each self-market question to the first matching category in the order restraint, outcome, effort. Restraint terms include quit, stop, avoid, abstain, stay off, sober, relapse, no-spend, no social media, and procrastinate. Outcome terms include accepted, admitted, hired, job offer, promoted, win, elected, approved, funded, pregnant, sick, COVID, pass an exam, score, rank, qualify, followers, and price. Effort terms include gym, work out, run, walk, lift, meditate, journal, write, post, publish, read, study, practice, finish, complete, submit, launch, build, code, clean, cook, apply, wake up, go to bed, streak, daily, and times per week. Questions matching none of the lists are unclassified. Many outcomes, such as exam results or audience growth, depend on both effort and outside forces; the categories are coarse.

\textbf{Encouragement.} A comment encourages the creator when it contains a well-wishing expression such as good luck, best of luck, you got this, you can do it, rooting for you, believe in you, go for it, keep going, fingers crossed, hope you make it, don't give up, or stay strong, or a flexed-biceps, four-leaf-clover, or crossed-fingers emoji. Congratulations (congrats, well done, good job, proud of you, and party or clapping emoji) are coded separately and not counted, because they can respond to an outcome that is already known; ambiguous cues (let's go, hearts) are in neither list. An observer counts as an encourager for a purchase only if the comment was posted before the purchase order was placed.

\textbf{Prior interaction.} An observer is a prior interactor of a creator in a market when the observer's first trade or comment in any other sampled market by the same creator predates the focal market's creation. Because the sample includes every self-market and up to \ExtCap{} external markets per self-market creator, but only one sampled market for most other creators, prior interaction is observed most completely for self-market creators.

\textbf{Orders and executions.} Price paths use every order, including cancelled and unfilled limit orders, in the order they were placed; an order's own price change is the difference between the probability after and before it. Purchases are fills with positive shares and positive cost, dated at the time of the fill, excluding the platform's automatic share redemptions; a cancelled order keeps the fills it received before cancellation. Positions count purchases and sales up to the order that set the final price; redemptions remove equal numbers of YES and NO shares and leave positions unchanged. A purchase is automated when it was placed through the platform's programming interface.

Table~\ref{tab:validation} compares these rules with the labels of two independent coders.

\input{generated/tab_validation}

\clearpage

\section{Additional tables}
\label{app:tables}

Tables~\ref{tab:gaps} and~\ref{tab:landmarks} give the detailed numbers behind Sections~\ref{sec:rq1} and~\ref{sec:rq2}. In both tables, a gap is the share of markets that resolved YES minus the price, in percentage points, so a negative gap means that prices overstated how often YES occurred.

\textbf{Gaps under other forecasts.} Table~\ref{tab:gaps} repeats the comparison of Section~\ref{sec:rq1} using prices other than the final price. Each row uses a different price as the forecast. The first three columns give the mean gap in each group of markets, and the last two give the difference between self-markets and each group of external markets, without controls. Three patterns stand out. First, in the upper rows, under every forecast other than the opening price, self-markets overstated YES more than external markets by other creators did, by \DiffGapThreeAbs{} to \DiffGapMidAbs{} pp. Second, the opening price is the exception. Most markets opened at the default 50\% and fewer than half resolved YES, so every group has a negative gap at the opening, but the gap was smaller for self-markets (\SelfGapOpen{} against \ExtGapOpen{} pp): before trading began, self-markets were not overpriced relative to external markets. Third, the bottom rows split the final price by range. The overstatement is concentrated in markets whose final price lay between 20\% and 80\% (difference \DiffRangeMidGap{} pp), whereas markets priced below 20\% or above 80\% were close to calibrated in both groups. Against the same creators' external markets the pattern is similar, but the differences for two early prices are smaller and not significant at the 5\% level (\DiffSameGapAday{} pp one day after creation and \DiffSameGapThree{} pp after the third observer purchase).

\input{generated/tab_gaps}

\textbf{Gaps over the market's life.} Table~\ref{tab:landmarks} follows calibration through the landmarks defined in Section~\ref{sec:measures}, from the opening to the final price; its last two columns are the estimates plotted in Figure~\ref{fig:calibration}B. For self-markets and external markets by other creators, each row reports how many markets were open at that point, their mean gap, and the YES rate implied for a market priced at 50\% (a value below 50\% means that such markets resolved YES less than half the time). The last two columns give the self-market difference at the same price level, for all markets and for markets in which the person never traded. Because these differences compare markets at the same price, they differ from the raw differences in Table~\ref{tab:gaps} (at the final price, \LmFinalDiff{} against \DiffGapFinal{} pp). A landmark a given time after creation and the landmark the same time before close cover the same markets, namely those that lasted at least that long, so the two 7-day rows compare the same markets early and late in their lives.

Reading down the table shows three patterns. At the opening, self-markets were, if anything, underpriced: at the same price they resolved YES \LmOpenDiffAbs{} pp more often than external markets. Within an hour the sign reversed, and the difference grew over the market's life, peaking at \LmWeekBeforeDiff{} pp a week before close before narrowing to \LmFinalDiff{} pp at the final price. The implied YES rate at 50\% shows the same shift: in self-markets it fell from \LmOpenSelfYesFifty{} at the opening to \LmWeekSelfYesFifty{} seven days after creation, whereas in external markets it stayed between \LmWeekExtYesFifty{} and \LmHourExtYesFifty{} in the first week and rose to \LmDayBeforeExtYesFifty{} a day before close. Finally, in markets in which the person never traded, the difference was larger at every landmark after the opening, from \LmHourNeverDiff{} pp after an hour to \LmWeekBeforeNeverDiff{} pp a week before close, so the overstatement arose even where only observers traded.

\input{generated/tab_landmarks}

%% file: generated/tab_flow.tex
\begin{table}[h]
\caption{Sample construction.}
\label{tab:flow}
\centering
\small
\begin{tabular}{lr}
\toprule
Step & Markets \\
\midrule
All markets retrieved from the platform & 194,347 \\
Binary (yes/no) markets & 140,090 \\
Denominated in mana & 139,818 \\
Resolved YES or NO & 98,456 \\
At least two unique traders & 94,701 \\
\quad of which first-person questions about the creator & 6,272 \\
\quad of which external markets & 81,791 \\
\quad of which other first-person markets & 6,638 \\
Sample: all first-person questions about the creator & 6,272 \\
Sample: external markets by self-market creators (up to 5 each) & 4,548 \\
Sample: external markets by other creators (random, quarter-stratified) & 6,272 \\
Sample total & 17,092 \\
\quad of which with at least one comment & 12,504 \\
\midrule
Analysis: forward-looking self-markets with an observer purchase & 6,150 \\
\quad first-person questions set aside (retrospective, normative, present-state) & 120 \\
Analysis: external markets by other creators with an observer purchase & 6,272 \\
Analysis: external markets by self-market creators with an observer purchase & 4,486 \\
\bottomrule
\end{tabular}
\par\smallskip
\begin{minipage}{0.85\textwidth}\footnotesize
Note: The analysis also sets aside 62 external markets whose creators asked only questions outside the forward-looking definition.
\end{minipage}
\end{table}

%% file: generated/tab_event.tex
\begin{table}[h]
\caption{Reconstruction of prices and executions from order records.}
\label{tab:event}
\centering
\small
\begin{tabular}{lr}
\toprule
Order records in the sample markets & 1,346,194 \\
Consecutive orders whose prices connect: all orders & 97.2\% \\
\quad filled, never-cancelled orders only & 85.9\% \\
Final price equal to the platform's reported probability: all orders & 99.4\% \\
\quad filled, never-cancelled orders only & 91.9\% \\
Cancelled orders with at least one fill & 141,685 \\
Purchase orders with a fill more than a minute after placement & 37,578 \\
Orders whose fills do not sum to the order's shares & 0 \\
Purchase executions before close & 901,706 \\
\quad from orders later cancelled & 17.0\% \\
\quad from limit orders & 30.2\% \\
\bottomrule
\end{tabular}
\par\smallskip
\begin{minipage}{0.85\textwidth}\footnotesize
Note: Two consecutive orders connect when the probability before the later order equals the probability after the earlier one. The reported probability is the market probability in the platform's market list.
\end{minipage}
\end{table}

%% file: generated/tab_resol.tex
\begin{table}[h]
\caption{Selection into a YES or NO resolution.}
\label{tab:resol}
\centering
\footnotesize
\setlength{\tabcolsep}{2.5pt}
\begin{tabular}{lrrrrrrrr}
\toprule
 & & \multicolumn{4}{c}{Status} & \multicolumn{3}{c}{Mean gap at last price (pp)} \\
\cmidrule(lr){3-6}\cmidrule(lr){7-9}
 & Markets & YES/NO & Cancelled & Probability & Unresolved & YES/NO only & Others as priced & Others NO; YES \\
\midrule
Self-markets & 6,475 & 92.0\% & 5.0\% & 1.3\% & 1.7\% & $-$4.5 & $-$4.1 & $-$7.9; $+$0.1 \\
External markets & 87,457 & 90.8\% & 4.6\% & 1.8\% & 2.8\% & $-$1.0 & $-$0.9 & $-$4.7; $+$4.6 \\
\bottomrule
\end{tabular}
\par\smallskip
\begin{minipage}{0.95\textwidth}\footnotesize
Note: Binary, mana-denominated markets with at least two traders whose close date was before 1 July 2026. ``Probability'' marks markets resolved to a probability. ``Others as priced'' counts every market without a YES or NO resolution with a gap of zero; the last column assigns all of them NO, or all of them YES, at their last price.
\end{minipage}
\end{table}

%% file: generated/tab_validation.tex
\begin{table}[t]
\caption{Validation of the coding rules against two coders' labels.}
\label{tab:validation}
\centering
\footnotesize
\setlength{\tabcolsep}{4pt}
\begin{tabular}{lrrr}
\toprule
 & Coder 1 & Coder 2 & Agreed labels \\
\midrule
\multicolumn{4}{l}{Agreement between the two coders, Cohen's $\kappa$ (percent agreement)} \\
\quad Question is about the creator & \multicolumn{3}{r}{0.98 (99.6\%)} \\
\quad Time of the event (future, present, past) & \multicolumn{3}{r}{0.93 (99.2\%)} \\
\quad Forecast rather than advice & \multicolumn{3}{r}{0.75 (99.0\%)} \\
\quad Creator would want YES & \multicolumn{3}{r}{0.91 (93.9\%)} \\
\quad Creator's control (effort, restraint, outcome, mixed) & \multicolumn{3}{r}{0.49 (64.7\%)} \\
\quad Outcome verifiable by others & \multicolumn{3}{r}{0.22 (54.8\%)} \\
\quad Comment encourages & \multicolumn{3}{r}{0.95 (98.0\%)} \\
\quad Comment congratulates & \multicolumn{3}{r}{0.95 (98.8\%)} \\
\quad Comment knows the outcome & \multicolumn{3}{r}{0.52 (84.2\%)} \\
\quad Comment doubts the creator & \multicolumn{3}{r}{0.84 (97.8\%)} \\
\midrule
\multicolumn{4}{l}{The paper's rules against the coders' labels} \\
\quad Random self-markets coded as forward-looking questions about the creator & 98.5\% & 99.5\% & 98.5\% \\
\quad Random external markets coded as not about the creator & 98.0\% & 99.0\% & 98.0\% \\
\quad Set-aside first-person questions coded as forward-looking & 20.0\% & 18.3\% & 15.0\% \\
\quad Unwanted-outcome rule: flagged questions coded unwanted (precision) & 90\% & 86\% & 90\% \\
\quad Unwanted-outcome rule: unwanted questions it flags (recall)$^a$ & 43\% & 47\% & 47\% \\
\quad Goal-type rule: category matches the coded control$^b$ & 69\% & 74\% & 66\% \\
\quad Encouragement rule: precision & 85\% & 89\% & 88\% \\
\quad Encouragement rule: recall$^a$ & 45\% & 43\% & 47\% \\
\quad Congratulation rule: precision & 98\% & 97\% & 98\% \\
\quad Congratulation rule: recall$^a$ & 50\% & 39\% & 66\% \\
\bottomrule
\end{tabular}
\par\smallskip
\begin{minipage}{0.95\textwidth}\footnotesize
Note: Two coders labelled 726 market questions and 400 comments independently, without seeing market outcomes or the rules' labels. Questions came from five strata: 200 random self-markets, every question the unwanted-outcome rule flags, the unflagged self-market questions containing an adverse word, 60 first-person questions set aside from the analysis, and 100 random external markets; comments were drawn from those flagged as encouraging, those flagged as congratulating, and the rest. ``Agreed labels'' uses the items on which the coders agreed. $^a$Weighted to the population of self-markets (questions) or comments by the sampling strata. $^b$Among random self-markets the rule classifies (127 of 200).
\end{minipage}
\end{table}

%% file: generated/tab_gaps.tex
\begin{table}[H]
\caption{Calibration gaps: share resolved YES minus price, in percentage points.}
\label{tab:gaps}
\centering
\footnotesize
\setlength{\tabcolsep}{4pt}
\begin{tabular}{lrrrrr}
\toprule
 & & \multicolumn{2}{c}{External markets} & \multicolumn{2}{c}{Self minus external} \\
\cmidrule(lr){3-4}\cmidrule(lr){5-6}
Price used as the forecast & Self-markets & Other creators & Same creators & Other creators & Same creators \\
\midrule
Final price & $-$4.4 (0.5) & $-$0.9 (0.3) & $-$1.3 (0.3) & $-$3.5 (0.6) & $-$3.1 (0.6) \\
Opening price & $-$4.8 (1.1) & $-$7.6 (0.7) & $-$9.7 (0.8) & $+$2.8 (1.3) & $+$4.9 (1.3) \\
Price one day after creation$^a$ & $-$9.2 (0.8) & $-$5.5 (0.6) & $-$7.4 (0.7) & $-$3.7 (1.0) & $-$1.8 (1.0) \\
Price one day before close$^a$ & $-$5.0 (0.7) & $-$0.2 (0.5) & $+$0.0 (0.5) & $-$4.8 (0.8) & $-$5.0 (0.8) \\
Price at mid-life & $-$7.5 (0.7) & $-$2.7 (0.5) & $-$3.7 (0.6) & $-$4.8 (0.9) & $-$3.9 (0.9) \\
Time-weighted average price & $-$7.0 (0.7) & $-$2.6 (0.5) & $-$3.5 (0.6) & $-$4.5 (0.8) & $-$3.5 (0.9) \\
Price after third observer purchase$^b$ & $-$8.3 (0.8) & $-$6.1 (0.6) & $-$7.5 (0.7) & $-$2.2 (1.0) & $-$0.8 (1.0) \\
\midrule
Final price, 20--80\% only & $-$9.4 (1.0) & $-$1.2 (1.6) & $-$3.2 (1.6) & $-$8.2 (1.9) & $-$6.2 (1.9) \\
Final price, $<$20\% or $>$80\% only & $-$1.5 (0.4) & $-$0.8 (0.2) & $-$0.8 (0.2) &  &  \\
\bottomrule
\end{tabular}
\par\smallskip
\begin{minipage}{0.95\textwidth}\footnotesize
Note: Mean of $Y-p$, where $Y=1$ if the market resolved YES; negative values mean that prices overstated how often YES occurred. ``Same creators'' are external markets made by the creators of self-markets. Differences are coefficients on a self-market indicator without other controls. Standard errors clustered by creator in parentheses. $^a$Markets open at the time (one day after creation: 5,383 self-markets and 5,437 external markets by other creators). $^b$Markets with at least three observer purchases.
\end{minipage}
\end{table}

%% file: generated/tab_landmarks.tex
\begin{table}[H]
\caption{Calibration at fixed points in the markets' lives.}
\label{tab:landmarks}
\centering
\footnotesize
\setlength{\tabcolsep}{3.5pt}
\begin{tabular}{lrrrrrrrr}
\toprule
 & \multicolumn{2}{c}{Markets open} & \multicolumn{2}{c}{Mean gap (pp)} & \multicolumn{2}{c}{YES rate at 50\%} & \multicolumn{2}{c}{Self minus external, same price (pp)} \\
\cmidrule(lr){2-3}\cmidrule(lr){4-5}\cmidrule(lr){6-7}\cmidrule(lr){8-9}
Point in the market's life & Self & External & Self & External & Self & External & All markets & Person never traded \\
\midrule
At the opening & 6,150 & 6,272 & $-$4.8 & $-$7.6 & 45\% & 42\% & $+$3.0 (1.3) & $-$1.2 (1.8) \\
1 hour after creation & 6,047 & 6,055 & $-$8.0 & $-$5.8 & 42\% & 43\% & $-$1.1 (1.1) & $-$3.9 (1.7) \\
6 hours after creation & 5,925 & 5,864 & $-$8.3 & $-$5.2 & 41\% & 43\% & $-$2.0 (1.1) & $-$5.5 (1.7) \\
1 day after creation & 5,383 & 5,437 & $-$9.2 & $-$5.5 & 40\% & 43\% & $-$2.4 (1.0) & $-$5.5 (1.7) \\
3 days after creation & 4,932 & 4,914 & $-$9.8 & $-$6.0 & 39\% & 42\% & $-$2.4 (1.1) & $-$5.7 (1.9) \\
7 days after creation & 4,204 & 4,356 & $-$10.7 & $-$6.5 & 38\% & 41\% & $-$2.9 (1.1) & $-$5.2 (1.9) \\
7 days before close & 4,204 & 4,356 & $-$6.9 & $-$1.3 & 39\% & 48\% & $-$5.3 (0.9) & $-$7.6 (1.5) \\
1 day before close & 5,383 & 5,437 & $-$5.0 & $-$0.2 & 42\% & 50\% & $-$4.6 (0.8) & $-$6.4 (1.3) \\
Final price & 6,150 & 6,272 & $-$4.4 & $-$0.9 & 39\% & 47\% & $-$2.4 (0.5) & $-$3.7 (1.0) \\
\bottomrule
\end{tabular}
\par\smallskip
\begin{minipage}{0.95\textwidth}\footnotesize
Note: External markets are those by other creators. A market enters a landmark after creation if it was still open at that time, and a landmark before the close if it had opened by then; the price at a landmark is the market probability after the last order at or before it. ``YES rate at 50\%'' is the implied share resolving YES for a market priced at one half, from a logistic recalibration. The same-price comparison regresses $Y-p$ on a self-market indicator with fixed effects for ten equal-width price bins; standard errors clustered by creator in parentheses. The last column keeps markets in which the creator made no trade before the final price (2,466 self, 2,538 external); this label uses the market's whole history.
\end{minipage}
\end{table}